\documentclass[graybox,envcountchap,sectrefs,natbib]{svmono}
\usepackage{mathptmx} 
\usepackage{helvet}       
\usepackage{courier}    
\usepackage{type1cm}  
\usepackage{amsfonts}
\usepackage{amssymb}
\usepackage{amsmath}
\usepackage{xspace}
\usepackage{chapterbib}
\usepackage{makeidx}         
\usepackage{multicol}        
\usepackage[bottom]{footmisc}
\usepackage{graphicx}        

\begin{document}
\setcounter{chapter}{8} %

%
%
\let\jnl=\rmfamily
\def\refe@jnl#1{{\jnl#1}}%

\newcommand\aj{\refe@jnl{AJ}}%
\newcommand\actaa{\refe@jnl{Acta Astron.}}%
\newcommand\araa{\refe@jnl{ARA\&A}}%
\newcommand\apj{\refe@jnl{ApJ}}%
\newcommand\apjl{\refe@jnl{ApJ}}%
\newcommand\apjs{\refe@jnl{ApJS}}%
\newcommand\ao{\refe@jnl{Appl.~Opt.}}%
\newcommand\apss{\refe@jnl{Ap\&SS}}%
\newcommand\aap{\refe@jnl{A\&A}}%
\newcommand\aapr{\refe@jnl{A\&A~Rev.}}%
\newcommand\aaps{\refe@jnl{A\&AS}}%
\newcommand\azh{\refe@jnl{AZh}}%
\newcommand\gca{\refe@jnl{GeoCh.Act}}%
\newcommand\grl{\refe@jnl{Geo.Res.Lett.}}%
\newcommand\jgr{\refe@jnl{J.Geoph.Res.}}%
\newcommand\memras{\refe@jnl{MmRAS}}%
\newcommand\jrasc{\refe@jnl{J.RoySocCan}}%
\newcommand\mnras{\refe@jnl{MNRAS}}%
\newcommand\na{\refe@jnl{New A}}%
\newcommand\nar{\refe@jnl{New A Rev.}}%
\newcommand\pra{\refe@jnl{Phys.~Rev.~A}}%
\newcommand\prb{\refe@jnl{Phys.~Rev.~B}}%
\newcommand\prc{\refe@jnl{Phys.~Rev.~C}}%
\newcommand\prd{\refe@jnl{Phys.~Rev.~D}}%
\newcommand\pre{\refe@jnl{Phys.~Rev.~E}}%
\newcommand\prl{\refe@jnl{Phys.~Rev.~Lett.}}%
\newcommand\pasa{\refe@jnl{PASA}}%
\newcommand\pasp{\refe@jnl{PASP}}%
\newcommand\pasj{\refe@jnl{PASJ}}%
\newcommand\skytel{\refe@jnl{S\&T}}%
\newcommand\solphys{\refe@jnl{Sol.~Phys.}}%
\newcommand\sovast{\refe@jnl{Soviet~Ast.}}%
\newcommand\ssr{\refe@jnl{Space~Sci.~Rev.}}%
\newcommand\nat{\refe@jnl{Nature}}%
\newcommand\iaucirc{\refe@jnl{IAU~Circ.}}%
\newcommand\aplett{\refe@jnl{Astrophys.~Lett.}}%
\newcommand\apspr{\refe@jnl{Astrophys.~Space~Phys.~Res.}}%
\newcommand\nphysa{\refe@jnl{Nucl.~Phys.~A}}%
\newcommand\physrep{\refe@jnl{Phys.~Rep.}}%
\newcommand\procspie{\refe@jnl{Proc.~SPIE}}%

\newcommand{\Al}{$^{26}$Al\xspace}
\newcommand{\al}{$^{26}$Al\xspace}
\newcommand{\Be}{$^{7}$Be\xspace}
\newcommand{\be}{$^{7}$Be\xspace}
\newcommand{\bem}{$^{10}$Be\xspace}
\newcommand{\ca}{$^{44}$Ca\xspace}
\newcommand{\Ca}{$^{44}$Ca\xspace}
\newcommand{\cam}{$^{41}$Ca\xspace}
\newcommand{\Co}{$^{56}$Co\xspace}
\newcommand{\co}{$^{56}$Co\xspace}
\newcommand{\csm}{$^{135}$Cs\xspace}
\newcommand{\ct}{$^{13}$C\xspace}
\newcommand{\ci}{$^{57}$Co\xspace}
\newcommand{\Ci}{$^{57}$Co\xspace}
\newcommand{\ch}{$^{60}$Co\xspace}
\newcommand{\Ch}{$^{60}$Co\xspace}
\newcommand{\Cl}{$^{36}$Cl\xspace}
\newcommand{\li}{$^{7}$Li\xspace}
\newcommand{\Li}{$^{7}$Li\xspace}
\newcommand{\Fe}{$^{60}$Fe\xspace}
\newcommand{\fh}{$^{60}$Fe\xspace}
\newcommand{\fe}{$^{56}$Fe\xspace}
\newcommand{\Fr}{$^{57}$Fe\xspace}
\newcommand{\fr}{$^{57}$Fe\xspace}
\newcommand{\mg}{$^{26}$Mg\xspace}
\newcommand{\Mg}{$^{26}$Mg\xspace}
\newcommand{\mn}{$^{54}$Mn\xspace}
\newcommand{\Na}{$^{22}$Na\xspace}
\newcommand{\Ne}{$^{22}$Ne\xspace}
\newcommand{\Ni}{$^{56}$Ni\xspace}
\newcommand{\nh}{$^{60}$Ni\xspace}
\newcommand{\Nh}{$^{60}$Ni\xspace}
\newcommand\nuk[2]{$\rm ^{\rm #2} #1$}  
\newcommand{\pd}{$^{107}$Pd\xspace}
\newcommand{\pb}{$^{205}$Pb\xspace}
\newcommand{\tc}{$^{99}$Tc\xspace}
\newcommand{\Sc}{$^{44}$Sc\xspace}
\newcommand{\Ti}{$^{44}$Ti\xspace}
\newcommand{\ti}{$^{44}$Ti\xspace}
\def\aa{$\alpha$}
\newcommand{\about}{$\simeq$}
\newcommand{\cms}{cm\ensuremath{^{-2}} s\ensuremath{^{-1}}\xspace}
\newcommand{\degree}{$^{\circ}$}
\newcommand{\flux}{ph~cm\ensuremath{^{-2}} s\ensuremath{^{-1}}\xspace}
\newcommand{\fluxrad}{ph~cm$^{-2}$s$^{-1}$rad$^{-1}$\ }
\newcommand{\ga}{\ensuremath{\gamma}}
\newcommand{\gam}{\ensuremath{\gamma}}
\def\nn{$\nu$}
\def\ra{$\rightarrow$}
\newcommand{\Msol}{M\ensuremath{_\odot}\xspace}
\newcommand{\msol}{M\ensuremath{_\odot}\xspace}
\newcommand{\Msolppc}{M\ensuremath{_\odot} pc\ensuremath{^{-2}}{\xspace}}
\newcommand{\Msolpy}{M\ensuremath{_\odot} y\ensuremath{^{-1}}{\xspace}}
\newcommand{\msb}{M\ensuremath{_\odot}\xspace}
\newcommand{\Msun}{M\ensuremath{_\odot}\xspace}
\newcommand{\Rsun}{R\ensuremath{_\odot}\xspace}
\newcommand{\rsun}{R\ensuremath{_\odot}\xspace}
\newcommand{\Lsun}{L\ensuremath{_\odot}\xspace}
\newcommand{\lsun}{L\ensuremath{_\odot}\xspace}
\newcommand{\solar}{\ensuremath{_\odot}\xspace}
\newcommand{\zs}{Z\ensuremath{_\odot}\xspace}



%
%
%

\chapauthor{Michael Wiescher\footnote{University of Notre Dame, Notre Dame, IN 46556, USA} and Thomas Rauscher\footnote{Universit\"at Basel, 4056 Basel, Switzerland}}
\chapter{Nuclear Reactions}
\label{sec:9:nuclearReactions}

Nuclear reaction rates determine the abundances of isotopes in
stellar burning processes. A multitude of reactions determine the
reaction flow pattern which is described in terms of reaction
network simulations. The reaction rates are determined by
laboratory experiments supplemented by nuclear reaction and
structure theory. We will discuss the experimental approach as well
as the theoretical tools for obtaining the stellar reaction rates.
A detailed analysis of a reaction is only possible for a few
selected cases which will be highlighted in this section. The bulk
of nuclear reaction processes is however described in terms of a
statistical model approach, which relies on global nuclear
structure and reaction parameters such as level density and mass
and barrier penetration, respectively. We will discuss a variety
of experimental facilities and techniques used in the field, this
includes low energy stable beam experiments, measurements at
radioactive beam accelerators, and neutron beam facilities.

\section{Nuclear Reactions in Astrophysical Environments}
\label{sec:9:astroreactions}

Nuclear reactions are the engine of stellar evolution and
determine the overall production of the long-lived radioactive
isotopes in a variety of nucleosynthesis patterns. A detailed
understanding of the characteristic production and depletion rates
within the framework of the nucleosynthesis process is crucial for
reliable model predictions and the interpretation of the observed
abundances.

There are several experimental and theoretical challenges in
obtaining stellar reaction rates.  The interaction energies in
stellar environments extend from basically zero projectile energy
up to only several MeV. This is especially challenging for the
measurement of the relevant reaction cross sections which can be
extremely small, especially for reactions with charged
projectiles. This also makes theoretical predictions extremely
difficult because several reaction mechanisms (see below) may
compete and simpler approximations may only be of limited use.
\index{Gamow window}

Another challenge arises from the fact that nuclear burning at
high temperature produces very short-lived isotopes which
subsequently decay to long-lived and stable species. Current
experimental technology can only access a fraction of those yet
and is still limited in obtaining detailed information on their
properties. The possibility to measure cross sections of reactions
with such highly unstable nuclei is even more limited currently.
Thus, investigations of nucleosynthesis in high-temperature
environments largely rely on theoretical models, which not only
have to treat the reaction mechanisms properly but also are
required to predict nuclear properties far from stability.

Moreover, depending on the actual plasma conditions, reactions in
an astrophysical plasma may proceed fundamentally differently from
those in the laboratory. This is due to two effects. On one hand,
laboratory nuclei are always part of atoms or molecules whereas
astrophysical nuclear burning involves fully ionized nuclei
immersed in a cloud of free electrons (and photons). The Coulomb
charge of a nucleus is partially shielded by the surrounding
electrons but this shielding (or screening) effect will be
different for an atom or molecule and a plasma because of the
different electron distribution and kinetics. While theoretical
cross sections always imply bare nuclei, the values have to be
appropriately converted (also based on a theoretical treatment of
different screening mechanisms) for comparison to low-energy
laboratory cross sections and for application in astrophysical
plasmas. Additionally, the quantum mechanically and geometrically
different electron distribution in a plasma directly affects
electron capture reactions. For example, nuclei such as $^7$Be or
$^{44}$Ti, decaying by capturing an electron from the K-shell of
the atom under laboratory conditions will not be able to do this
in a stellar plasma. Instead, electron capture inside a star
involves capturing a free electron from the plasma, which is more
unlikely and therefore the terrestrial half-life can be shorter
than the one in a stellar environment
\citep{sec9:Il07,sec9:JKKL92}.

Finally, due to the high photon and matter densities in
astrophysical environments, nuclei very quickly reach thermal
equilibrium with the surroundings by excitation and de-excitation
via photons and by collisions. In most cases, this happens on a
shorter timescale than that of nuclear transformations (one of the
exceptions being isomeric states). Consequently, the nuclei
involved in the reactions occur not only in their ground states,
as in the laboratory, but also their excited states are populated
with a probability involving the Boltzmann factor.
\index{Maxwell-Boltzmann distribution}
So far,
reactions on excited states can only be treated theoretically. The
population of excited nuclear states does not only depend on the
plasma temperature but also on the structure of a nucleus. Nuclei
with isolated levels reachable within a few keV projectile energy
will exhibit pronounced thermal population already at low plasma
temperature. This is especially important for modern s-process
studies which require high accuracy knowledge of neutron capture
rates. Thermal effects are also important in decays and neutrino
reactions because the available phase space of the reaction
products is altered, leading to a modification of the rate. For
example, electron capture rates in the stellar core collapse are
enhanced at temperatures $T>1.5$ MeV because of the unblocking of
low-lying neutron states by thermal excitation \citep{sec9:CW84}.

\subsection{Reaction Networks and Thermonuclear Reaction Rates}
\label{sec:9:net_rate}

The change of abundances $Y$ with time due to nuclear processes is
traced by coupled differential equations. To be fully solvable,
the number of equations $N$ must equal the number of involved
nuclei acting as reaction partners and thus an equation matrix of size $N^2$ has
to be solved. Such a set of coupled
equations is called reaction network and can generally be written
as
\begin{equation}
\label{eq:9:network}
 \dot{Y}_i=\frac{1}{\rho N_\mathrm{A}} \dot{n}_i = \frac{1}{\rho N_\mathrm{A}} \left\{ \sum_j {^{1}_{i}P_{j}} \; {_{i}\lambda_{j}} + \sum_{j} {^{2}_{i}P_{j}}\; {_{i}r_{j}} +
 \sum_j {^{3}_{i}P_{j}}\; {_{i}\hat{r}_{j}} + \dots \right\} \quad,
\end{equation}
where $1\leq i\leq N$ numbers the nucleus, $_i\lambda_j$ is the
$j$th rate for destruction or creation of the $i$th nucleus
without a nuclear projectile involved (this includes spontaneous
decay, lepton capture, photodisintegration), and $_ir_j$ is the
rate of the $j$th reaction involving a nuclear projectile and creating or
destroying nucleus $i$. Similarly, we have three-body reactions where nucleus $i$
is produced or destroyed together with two other (or similar) nuclei.
Reactions with more participants (denoted by $\dots$
above) are unlikely to occur at astrophysical conditions and are
usually neglected. The quantities $^1 _iP_j$, $^2
_iP_j$, and $^3 _iP_{jk}$
are positive or negative integer numbers specifying the amount of
nuclei $i$ produced or destroyed, respectively, in the given process. As shown
below, the rates $\lambda$, $r$, and $\hat{r}$ contain the abundances of the
interacting nuclei. Rates of type $\lambda$ depend on one abundance (or number density),
rates $r$ depend on the abundances of two species, and rates $\hat{r}$ on three.

Using abundances $Y$ instead of number densities $n=Y \rho N_\mathrm{A}$ (where $\rho$ is the plasma density)
has the advantage that a change in the number of nuclei in a given volume due to density fluctuations is
factored out and only changes by nuclear processes are considered.
Using abundance changes, the total energy generation rate per mass due to
nuclear reactions can easily be expressed as
\begin{equation}
\dot \epsilon  = - \sum_i \dot Y_i N_A M_{i}c^2 \quad,
\end{equation}
with the rest masses $M_{i}c^2$ of the participating nuclei.

The rates $_i\lambda_j$ appearing in the first term of Eq.\
\ref{eq:9:network} are reactions per time and volume, and only contain the
abundance $Y_j$. For example, ${_{i}\lambda_{j}}$ is
simply $n_j L_j=Y_j \rho N_\mathrm{A} L_j$ for $\beta$-decays. The factor
$L_j=(\ln 2) / ^jT_{1/2}$ is the usual decay constant (with the unit 1/time) and is related to the half-life $^jT_{1/2}$ of the
decaying nucleus $j$. It has to be noted that some decays depend on the plasma temperature and thus
$L_j$ is not always constant, even for decays.

Two-body rates $r$ include the abundances of two interacting particles or nuclei.
In general, target and projectile follow specific thermal
momentum distributions $dn_1$  and $dn_2$ in an astrophysical plasma.
With the resulting relative velocities $\vec v_1 -\vec v_2$,
the number of reactions per volume and time, is given by
\begin{equation}
\label{eq:9:2body}
r_{12}=\int \hat{\sigma}(\vert \vec v_1 -\vec v_2\vert)
 \vert \vec v_1 -\vec v_2\vert dn_1 dn_2 \quad,
\end{equation}
and involves the reaction cross section $\hat{\sigma}$ as a function of velocity/energy,
the relative velocity $\vec v_1 -\vec v_2$ and the thermodynamic distributions
of target and projectile $dn_1$ and $dn_2$.
The evaluation of this integral depends on the type of particles
(fermions, bosons) and distributions which are involved.

However, many two-body reactions can be simplified and effectively expressed similarly to
one-body reactions, only depending on one abundance (or number density).
If reaction partner 2 is a photon, the relative velocity is always $c$ and the quantities in the integral do not depend on $dn_1$.
This simplifies the rate expression to
\begin{equation}
\lambda_1=L_{\gamma}(T) n_1\quad,
\end{equation}
where $L_{\gamma}(T)$ stems from an integration over a
Planck distribution for photons of temperature $T$. This is similar to the decay rates introduced earlier and therefore we replaced $r$ by $\lambda$ in our notation and can include this type of reaction in the first term of Eq.\ \ref{eq:9:network}.
A similar procedure is used for electron captures by protons and nuclei.
Because the electron is about 2000 times less massive than a nucleon, the
velocity of the nucleus is negligible in the center-of-mass system in
comparison to the electron velocity ($\vert \vec v_\mathrm{nucleus}- \vec v_\mathrm{electron} \vert
\approx \vert \vec v_\mathrm{electron} \vert$). The electron
capture cross section has to be integrated over a
Fermi distribution of electrons. \index{Fermi!energy} \index{Fermi!distribution}
The electron capture rates are a function of $T$ and $n_e=Y_e \rho N_A$, the
electron number density. In a neutral, completely
ionized plasma, the electron abundance $Y_e$ is equal to the total proton
abundance $Y_e=\sum_i Z_i Y_i$ and thus
\begin{equation}
\lambda_\mathrm{nucleus,ec}=L_\mathrm{ec} (T,\rho Y_e)n_\mathrm{nucleus} \quad.
\end{equation}
Again, we have effectively a rate per target $L$ (with unit 1/time)
similar to the treatment of decays earlier and
a rate per volume including the number density of only one nucleus. We denote the latter
by $\lambda$ and use it in the first term of Eq.\ \ref{eq:9:network}.
This treatment can be applied also to the capture of positrons,
being in thermal equilibrium with photons, electrons, and nuclei.
Furthermore, at high densities ($\rho >10^{12}$gcm$^{-3}$) the size of the neutrino
scattering cross section on nucleons, nuclei, and electrons ensures that enough
scattering events occur to lead to a continuous neutrino energy distribution.
\index{neutrino!energy}
Then also the inverse
process to electron capture (neutrino capture) can occur as well as
other processes like, e.g., inelastic scattering, leaving a nucleus in an excited
state which can emit nucleons and $\alpha$ particles.
Such reactions can be expressed similarly to photon and electron
captures, integrating over the corresponding neutrino distribution.

In the following, we focus on the case of two interacting nuclei or nucleons as these
reactions will be extensively discussed in Secs.\ \ref{sec:9:reactionmodels} and
\ref{sec:9:expfac}. This will result in
an actual two-body rate $r$ to be used in the second term of Eq.\ \ref{eq:9:network}. Here, we mention in passing that Eq.\ \ref{eq:9:2body} can be generalized to 3 and more interacting nuclear species by integrating over the appropriate number of distributions, leading to rates $\hat{r}$ and higher order terms in Eq.\ \ref{eq:9:network}.

Turning our attention back to two-body reactions, we note that the velocity distributions can be replaced by energy distributions. Furthermore, it can be shown that
the two distributions in Eq.\ \ref{eq:9:2body} can be replaced by a single one in the center-of-mass system. This time the resulting expression describes a rate $r$ including two abundances (or number densities) and showing up in the second term of Eq.\ \ref{eq:9:network}. The rate $r$ is defined as an
interaction of two reaction partners with an energy distribution
$\phi(E)$ according to the plasma temperature $T$ and a reaction
cross section $\sigma (E)$ specifying the probability of the
reaction in the plasma:
\begin{equation}
r=\frac{n_1 n_2}{1+\delta_{12}} \int_0^\infty \sigma (E) \phi (E)\,dE \quad.
\end{equation}
The factor $1/(1+\delta_{12})$ with the Kronecker symbol $\delta$
is introduced to avoid double counting. The nuclear cross section
is defined as in standard scattering theory by
\begin{equation}
\sigma = \frac{\mathrm{number\ of\ reactions\ target}^{-1} \mathrm{sec}^{-1}}
{\mathrm{flux\ of\ incoming\ projectiles}}\quad.
\end{equation}
However, in an astrophysical plasma, nuclei quickly (on the
timescale of nuclear reactions and scattering) reach thermal
equilibrium with all plasma components. This allows thermal
excitation of nuclei which follows a Boltzmann law and gives rise
to the {\it stellar} reaction rate
\begin{align}
\label{sec9:eq:csstar}
r^*=&\frac{n_1 n_2}{1+\delta_{12}}\frac{1}{\sum_x (2J_x+1) e^{-\frac{E_x}{kT}}}
\sum_x \left\{(2J_x+1) \int_0^\infty \sigma^x(E^x) \phi (E^x) e^{-\frac{E_x}{kT}} dE^x\right\}
\nonumber \\=&\frac{n_1 n_2}{1+\delta_{12}}\frac{1}{G(T)}\sum_x \left\{(2J_x+1) \int_0^\infty \sigma^x(E^x) \phi (E^x) e^{-\frac{E_x}{kT}} dE^x\right\}\quad,
\end{align}
where the sum runs over all excited states $x$ of the target (for
simplicity, here we assume the projectile, i.e.\ the second
reaction partner, does not have excited states) with spin $J_x$
and excitation energy $E_x$. The quantity $G$ is the partition
function of the nucleus. The cross section $\sigma^x$ includes the reactions commencing from excited state $x$
and they are functions of the energy $E^x$ relative to this excited state.
Cross sections $\sigma=\sigma^{x=0}$
measured in terrestrial laboratories do not include such thermal
effects. At low temperature (e.g., for the s-process) the stellar
enhancement factor $\mathrm{SEF}=r^*/r$ will only differ
from unity when there are excited states within a few keV above
the reaction threshold. At the large temperatures reached in
explosive burning, thermal enhancement can lead to a considerable
deviation from the ground-state cross section, see also \citet{raureview}.
\index{nucleosynthesis!explosive} \index{Maxwell-Boltzmann distribution}

Nuclei in an astrophysical plasma obey a Maxwell-Boltzmann (MB) distribution $\phi(E) = \phi_\mathrm{MB}$ and we obtain finally \citep{raureview}:
\begin{equation}
r  =  \frac{n_1 n_2}{1+\delta_{12}} <\sigma v>^*  \quad, \end{equation} \begin{equation}
<\sigma v>^*=(\frac{8}{\mu \pi})^{1/2}
(kT)^{-3/2} \frac{1}{G(T)}\sum_x \left\{(2J_x+1) \int_0 ^\infty {E^x \sigma^x (E^x) e^{-\frac{E^x+E_x}{kT}}\,dE^x}\right\} \quad.\label{sec9:eq:rate}
\end{equation}
Here, $\mu$ denotes the reduced mass of the two-particle system
and $<\sigma v>^*$ is the stellar reaction rate per particle pair or {\it
reactivity}.

As mentioned above the charge of the reaction partners can be
screened. For most astrophysical conditions this can be included
by introducing a screening factor $f_\mathrm{screen}$, modifying
the above rate for bare nuclei \citep{sec9:Il07,sec9:SvH69}
\begin{equation}
r^\mathrm{scr}=f_\mathrm{screen} r\quad.
\end{equation}
The screening factor is derived from the plasma conditions of the
specific stellar environment. At high densities and low
temperatures screening factors can enhance reactions by many
orders of magnitude and lead to {\it pycnonuclear ignition}
\citep{sec9:YGA06}. However, note that the above factorization is
not valid for vanishing temperatures when nuclei are trapped in a
Coulomb lattice.

Forward and reverse rates are related. Applying the well-known
reciprocity theorem for nuclear transitions \citep{sec9:BW52} and
further assuming that the reactands in the entrance channel $a$ as
well as the reaction products in the exit channel $b$ are
instantaneously thermalized (the {\it detailed balance}
principle), the relation \citep{sec9:HWFZ76,sec9:Il07}
\begin{equation}
\label{sec9:eq:detbal}
<\sigma v>^*_{b\rightarrow a} = \frac{1+\delta_{b_1 b_2}}{1+\delta_{a_1 a_2}} \frac{G_{a_1}G_{a_2}}{G_{b_1}G_{b_2}}
\left( \frac{\mu_a}{\mu_b} \right) ^{3/2} e^{-\frac{Q}{kT}} <\sigma v>^*_{a\rightarrow b} \quad,
\end{equation}
relating the {\it stellar} reverse rate to the {\it stellar}
forward rate. The latter has the reaction $Q$-value $Q$.
 For captures (forward channel $a$) and photodisintegrations
(reverse channel $b$), Eq.\ \ref{sec9:eq:detbal} transforms to
\begin{equation}
\label{sec9:eq:detbalgam}
L_\gamma=\frac{1}{1+\delta_{a_1 a_2}} \frac{G_{a_1}G_{a_2}}{G_{b}}
\left( \frac{\mu_a kT}{2\pi\hbar^2} \right) ^{3/2} e^{-\frac{Q}{kT}} <\sigma v>^*_\mathrm{capture} \quad .
\end{equation}
These expressions will not be valid anymore if any of the involved
rates was derived from a laboratory cross section. They also imply
that the detailed balance assumption is valid. Detailed balance
can be violated in nuclei with long-lived isomeric states which
are not populated or depopulated during regular reaction
timescales. For these cases, reactions to separate final states
have to be calculated and the (de)population of these states by
photon transitions followed explicitly \citep{sec9:WF80}.
Important examples for such nuclei are $^{26}$Al and $^{180}$Ta
\citep{sec9:rhhw92}.

\subsection{Reaction Equilibria}
\label{sec:9:equil} It is not always necessary to solve a full
reaction network (Eq.\ \ref{eq:9:network}) including all the
rates. On one hand, simplifications can often be made by omitting
slow reactions which will not contribute significantly during the
timescale of the astrophysical event. These are, for example,
charged-particle reactions on heavy targets in hydrostatic stellar
burning. On the other hand, high temperature can establish
reaction equilibria. When both forward and reverse reactions
become sufficiently fast to reach equilibrium with abundances set
at equilibrium values. The equilibrium abundances of nuclei can be
derived by using the relations \ref{sec9:eq:detbal} and
\ref{sec9:eq:detbalgam} in the network equation
\ref{eq:9:network} and assuming $\dot{Y}=0$. Somewhat depending
on the density, for $T>4-5$ GK all reactions are in a full {\it
nuclear statistical equilibrium} (NSE) and the abundances are
\index{process!nuclear statistical equilibrium}
given by
\begin{equation}
\label{sec9:eq:nse}
Y_i=G_i \left( \rho N_\mathrm{A} \right) ^{A_i-1} \frac{A_i^{3/2}}{2^A_i} \left( \frac{2\pi \hbar ^2}{m_\mathrm{u}kT} \right) ^{\frac{3}{2}(A_i-1)} e^\frac{B_i}{kT} Y_\mathrm{n}^{N_i} Y_\mathrm{p}^{Z_i} \quad,\\
\end{equation} \begin{equation}
 \sum_i A_i Y_i=1
\sum_i Z_i Y_i=Y_\mathrm{e} \end{equation}
with $Z_i$, $N_i$, $A_i$, and $B_i$ being the charge, neutron number, mass number, and
the binding energy of the nucleus $i$, respectively, the atomic
mass unit $m_\mathrm{u}$, and the abundances of free neutrons
$Y_\mathrm{n}$, free protons $Y_\mathrm{p}$, and free electrons
$Y_\mathrm{e}$. Here, it is assumed that reactions via the strong
and electromagnetic interactions are in equilibrium while the weak
interaction is not. Therefore, $Y_\mathrm{e}$ can still be
time-dependent and thus also the resulting NSE abundances $Y_i$.

At $T<4$ GK and/or low densities only some reactions may be in
equilibrium while others are too slow. This gives rise to the
so-called {\it quasi-statistical equilibrium} (QSE) where only
groups of nuclei are equilibrated and those groups are connected
by slower reactions which are not in equilibrium
\citep{sec9:HT99}. Abundance ratios within a QSE group can be
determined by application of Eq.\ \ref{sec9:eq:nse} while the
connecting, slow reaction determines the amount of matter in each
group relative to the other groups at a given time. QSE occurs in
low temperature, low density Si-burning and in O-burning of stars.
Often, the slowest rate falling out of equilibrium first is that
of the strongly density-dependent triple-$\alpha$ reaction.

A special case of a QSE is the {\it waiting-point approximation},
\index{rocess!quasiequilibrum} \index{waiting points} \index{process!r process}
often used in r-process calculations
\citep{sec9:fiss1,sec9:fiss2}. There, the network is reduced to
neutron capture reactions and their reverse reactions, and
$\beta^-$-decay (with possible release of neutrons). Assuming
equilibrated capture and photodisintegration, QSE within an
isotopic chain is obtained and the relative abundances are given
by
\begin{equation}
\frac{Y_{i+1}}{Y_i}=n_\mathrm{n} \frac{G_{i+1}}{2G_i} \left( \frac{A_i+1}{A_i} \right) ^{3/2} \left( \frac{2\pi \hbar ^2}{m_\mathrm{u} kT} \right) ^{3/2} e^\frac{Q_\mathrm{ncap}}{kT} \quad.
\end{equation}
The neutron number density is denoted by $n_\mathrm{n}$ and the
neutron capture $Q$-value is given by the neutron separation
energy in nucleus $i+1$: $Q_\mathrm{ncap}=S_{\mathrm{n},i+1}$. The
indices $i$ are ordered by increasing neutron number. The
$\beta^-$-decays are much slower and not in equilibrium. Synthesis
of the next element is delayed until the decay of the produced
isotopes. Typically, only one or two nuclides have significant
abundances in such an isotopic QSE chain, hence the name
waiting-point approximation.

The advantage of using equilibria is that the rates -- and thus
the cross sections -- do not have to be known explicitly. The
resulting abundances are completely determined by basic nuclear
properties and the conditions in the astrophysical environment.

\section{Relevant Energy Range of Astrophysical Cross Sections}

In the general calculation of the reaction rate according to Eq.\
\ref{sec9:eq:rate} the nuclear cross section has to be known.
Although the integration limits in Eq.\ \ref{sec9:eq:rate} run
from Zero to Infinity, significant contributions to the integral
only come from a comparatively narrow energy range. This is due to
the shape of the MB distribution, showing a peak around the energy
$E_\mathrm{MB}=kT$ and quickly approaching very small values both
towards $E=0$ and $E\gg kT$. For a slowly varying cross section
(as found, e.g., in non-resonant neutron-induced reactions), the
relevant energy range is simply given by the peak of the
distribution, $E_0=E_\mathrm{MB}$ and its width
$\Delta_0=\Delta_\mathrm{MB}$. For partial waves higher than
s-waves, the additional centrifugal barrier introduces a stronger
energy dependence in the cross section and shifts the relevant
range to slightly higher energy, i.e.\ $E_0\approx 0.172 T_9 (\ell
+ 1/2)$ MeV and $\Delta_0 \approx 0.194 T_9 \sqrt{\ell + 1/2}$ for
partial waves $\ell > 0$ \citep{sec9:RTK97,sec9:Wag69}.
Charged-particle cross sections exhibit a strong energy dependence
at energies close to and below the Coulomb barrier. They decrease
by many orders of magnitude towards lower energy. Using the
astrophysical S-factor
\begin{equation}
\label{sec9:eq:sfactor}
 S(E)=\sigma E e^{2\pi \eta} \quad,
\end{equation}
with $\eta$ being the Sommerfeld parameter describing the barrier
penetrability, most of the Coulomb suppression is taken out and
$S(E)$ is easier to handle because it is varying less with energy
than $\sigma$. Inserting definition \ref{sec9:eq:sfactor} into
Eq.\ \ref{sec9:eq:rate} shows that the penetration factor causes
a significant shift of the relevant energy range towards higher
energy. The resulting energy window (the {\it Gamow window}, given
\index{Gamow window}
by the {\it Gamow peak} appearing when folding the charged
particle cross section with the MB distribution) can be
approximated by \citet{sec9:Il07,sec9:RTK97}
\begin{equation}
 \label{sec9:eq:gamow}
 E_0=0.12204 \left(Z_1^2 Z_2^2 \mu \right) ^{1/3} T_9 ^{2/3} \quad \mathrm{MeV} \end{equation} \begin{equation}
 \Delta_0=4\sqrt{\frac{E_0kT}{3}}=0.23682 \left(Z_1^2 Z_2^2 \mu \right) ^{1/6} T_9 ^{5/6} \quad \mathrm{MeV}\quad .
\end{equation}
Here, $T_9$ is the plasma temperature in GK. The idea of a single,
relevant energy window is only viable for non-resonant cross
sections or reactions with broad resonances. Strong, narrow
resonances lead to fragmentation of the peak and split it up in
several small energy ranges around the resonance energies, with
decreasing weight towards higher energy.

It is important to note that Eq.\ \ref{sec9:eq:gamow} is not always
valid. It is based on the assumption that the energy dependence of
the cross section is mainly determined by the penetration of the
projectile through the Coulomb barrier. However, the dependence is
dominated by the one of the smallest width in the entrance or exit channel
for resonant reactions or smallest \textit{averaged} width in the case of Hauser-Feshbach
compound reactions (see Sec.\ \ref{sec:9:highdensitymodels}). This smallest width
can also be the one of the exit channel, leading to a different maximum
in the contribution to the reaction rate integral than estimated from Eq.\ \ref{sec9:eq:gamow}.
This is often the case in capture reactions when
$\Gamma_\mathrm{projectile} \gg \Gamma_\gamma$ \citep{sec9:Il07,sec9:NICCPU07}. Because of the weak energy
dependence of the $\gamma$-width, there would not be a Gamow window. Effectively, however,
the Gamow window is shifted to energies where $\Gamma_\mathrm{projectile}$ (which is strongly
energy dependent) becomes smaller than $\Gamma_\gamma$. Since reaction rates at higher temperature
are determined by cross sections at higher energy, the discrepancy between Eq.\ \ref{sec9:eq:gamow}
and the true maximum of the integrand is more pronounced at high temperature than at low temperature. Therefore, the relevant energy range for reactions in explosive burning should be
derived by a proper inspection of the product of the (predicted) cross sections and the MB distribution (see \citet{rauenergies} for details). For other charged particle captures in astrophysics,
 often $\Gamma_\mathrm{projectile} \ll \Gamma_\gamma$ due to the low interaction energy implied by
$E_\mathrm{MB}=kT=T_9 /11.6045$ MeV, unless for light nuclei (with
low Coulomb barrier). Regarding neutron captures, although $\Gamma_\mathrm{n} \gg \Gamma_\gamma$ will apply in most cases (unless very close to the reaction
threshold), the shape of the integrand is mostly determined by the shape of the MB distribution and obviously not by any Coulomb penetration.
Therefore, the relevant energy window for neutrons can still be estimated from the MB distribution
as shown above.

\section{Nuclear Reaction Models}
\label{sec:9:reactionmodels}

Having determined the relevant energy range, the cross sections
have to be predicted by reaction models or determined
experimentally. As previously mentioned, often measurements for
astrophysics prove difficult due to small cross sections or/and
unstable nuclei involved. However, even if a measurement is
feasible, the resulting cross section has to be corrected for
effects of electron screening and thermal excitation of the target
via theoretical models before being used to compute an
astrophysical reaction rate.

Here, we provide a brief overview of approaches to predict
low-energy cross sections of reactions involving the strong force.
Decays and other reactions via the weak force are important but
cannot be discussed due to limited space. The reader is referred
to other sources, e.g. \citep{sec9:weak1,sec9:weak2,sec9:weak3}
and references therein. We also do not cover fission reactions
which are important in extremely neutron-rich explosive
environments where a r-process could occur and reach the region of
fissionable nuclei. Current predictions of fission barriers,
however, carry large uncertainties. For details, see e.g.
\citep{sec9:fiss1,sec9:fiss2,sec9:fiss3,sec9:fissgor,sec9:fissigornew} and references therein.
We also only discuss reactions between a nucleus and a nucleon or
an $\alpha$-particle as the majority of reactions in astrophysics
is of that type.

The interaction of a particle with a nucleus can excite few or
many degrees of freedom, i.e. transfer energy to few (or none) or
to many of the nucleons constituting the target nucleus. In
nature, all interaction types are, in principle, possible but
often only one will be dominating at a given interaction energy
but with gradual transitions from one type to the other within
certain energy intervals. For theory, it is simpler to consider
extreme, idealized cases. Interdependence and interference effects
between different reaction mechanisms, even if in principle
understood, are very difficult to predict and especially so for
the required large number of reactions with unstable nuclei
required in astrophysics. In the following we introduce a
selection of relevant reaction mechanisms considered in
literature. The number of degrees of freedom which can be excited
depends on the number of states or levels present in the system
formed by projectile and target \citep{sec9:DR06}. Therefore, it
is helpful to distinguish between compound systems with low and
high level densities.

\subsection{Resonance and Potential Models}
\label{sec:9:lowdensitymodels}

Low level-density systems exhibit no or only few, isolated
resonances in the relevant energy range. These involve mostly
light nuclei which have few, widely spaced excited states within
several tens of MeV above the ground state and therefore also show
only few resonances even when the separation energy of the
projectile from the compound system is large. A similar situation
also occurs for heavier nuclei with closed shells or heavier
nuclei far off stability and close to the driplines where the
projectile separation energy becomes very low (e.g. in neutron
capture on extremely neutron-rich nuclei) and in consequence the
compound system is formed at very low relative energy.

In principle, isolated resonances can be included by the {\it
Breit-Wigner formula} \citep{sec9:BW52}
\begin{equation}
 \sigma^x= \frac{\omega ^2}{4\pi} \frac{2J+1}{(2J_x+1)(2J_\mathrm{proj}+1)} \frac{\Gamma^x_a \Gamma_b}{\left(E-E_\mathrm{res}\right)^2+\frac{\Gamma_\mathrm{tot}^2}{4}} \quad,
\end{equation}
where $J$ and $E_\mathrm{res}$ refer to the spin and energy of the
resonance, $\omega$ is the de Broglie wavelength, and
$\Gamma_\mathrm{tot}$ is the total resonance width, including the
entrance and exit widths $\Gamma^x_a$ and $\Gamma_b$ plus all
other open channels. Note that the widths are energy dependent.
For a narrow resonance, inserting the above in Eq.\
\ref{sec9:eq:rate} yields
\begin{equation}
 N_\mathrm{A}\left< \sigma v \right>=1.54\times 10^{11} \frac{1}{(\mu T_9)^{3/2}} \frac{2J+1}{(2J_x+1)(2J_\mathrm{proj}+1)} \frac{\Gamma^x_a \Gamma_b}{\Gamma_\mathrm{tot}}
e^{-\frac{11.6045 E_\mathrm{res}}{T_9}} \quad .
\end{equation}
This gives the reactivity in units of $\mathrm{cm}^3
\mathrm{s}^{-1} \mathrm{mole}^{-1}$ when the widths and the
resonance energy $E_\mathrm{res}$ are inserted in units of MeV
\citep{sec9:Il07}. (Note that the above equations do not involve
stellar cross sections. For a true stellar cross section and rate,
a thermally weighted sum of target states has to be used,
according to Eq.\ \ref{sec9:eq:csstar}.) However, tails of
resonances with the same $J$ may interfere and there may also be
interference with a direct component (see below). Therefore,
additional interference terms may have to be added (see, e.g.,
\citep{sec9:RR96}). Furthermore, location of the resonance and the
widths have to be predicted from nuclear structure. Currently,
this is not possible from first principles (except for the
lightest nuclei) with the accuracy needed in applications.
Therefore, this information usually has to come from experiments.

Instead of Breit-Wigner formulas and interference terms, often the
{\it R-matrix method} \citep{sec9:LT58} is used. It is applied to
parameterize experimentally known cross sections with as few
parameters as possible, implicitly accounting for resonances and
their interference. The R-matrix approach can be used to
extrapolate the nuclear cross section from existing data to the
Gamow range as long as nuclear structure information about
resonance levels and non-resonant reaction contributions are
included.

In addition to possible resonance contributions a direct capture
process can occur. These are fast, one-step processes in which a
captured particle directly enters the final state. Typical
reaction timescales of direct processes are of the order of
10$^{-22}$ s whereas compound reactions, distributing the energy
among a large number of nucleons, take of the order of 10$^{-16}$
s. Direct reactions include transfer processes where a particle
exchange takes place between projectile and target nucleus, and
capture processes in which the projectile is being fully captured
by the target nucleus. These two reaction types can be treated in
ab initio models, determining the cross sections from wave
functions obtained by solving the Schr\"odinger equation using
effective potentials.

For transfer reactions often the \textit{Distorted Wave Born
Approximation} (DWBA) \citep{sec9:Sat,sec9:Glen} is used,
utilizing optical potentials to compute the cross sections from
the overlap integral of distorted scattering wave functions and
the bound state wave function. The DWBA implicitly assumes that
elastic scattering is dominant while non-elastic contributions can
be treated perturbatively.

On the other hand, capture reactions can be calculated with a
simple potential model, which is a first-order approach involving
an electromagnetic operator describing the emission of photons due
to the dynamics in the movement of electric charges
\citep{sec9:DR06}. In the potential model the differential cross
section is proportional to the matrix element defined by the
overlap of the final state $\phi_\beta$ of the final nucleus and
the initial state composed of the target wave function
$\phi_\alpha$ and a (distorted) scattering wave of the projectile
$\chi_\alpha$. This can be decomposed into an overlap function $S$
of the target and the final nucleus and a radial integral
containing the scattered wave $\chi^x_\alpha (r)$, the bound state
wave function of the projectile in the target $\phi_\mathrm{a+A}$,
and the radial form of the electromagnetic operator
$\mathcal{O}_\mathrm{EM}$ \citep{sec9:dc}
\begin{equation}
 \frac{d\sigma^x}{d\Omega} \propto \left| \left< \phi_\beta | \mathcal{O}_\mathrm{EM} | \chi_\alpha \phi^x_\alpha \right> \right|^2 \propto S \left| \phi_\mathrm{a+A} (r) \mathcal{O}_\mathrm{EM}(r) \chi^x_\alpha (r) \,dr \right|^2 \quad.
\end{equation}
The wave functions $\phi_\mathrm{a+A} (r)$ and $\chi^x_\alpha (r)$
 are obtained by solving the radial Schr\"odinger equation with
appropriate effective potentials.

Both approaches, DWBA and potential model, require a
renormalization of the resulting cross section through
spectroscopic factors $S$, describing nuclear structure effects by
the overlap of initial and final state of the system. These
spectroscopic factors have to be obtained from nuclear structure
models or by comparison with experiment
\citep{sec9:Sat,sec9:Glen}.

Microscopic reaction models are first principle methods, starting
from effective nucleon-nucleon interactions and treating all
nucleons in a Hamiltonian with exact antisymmetrization of the
wave functions. Because of this, no artificial distinction between
direct and resonant contributions has to be made. Unfortunately,
such reaction models are limited to systems of few nucleons.
Although the Quantum Monte Carlo method \citep{sec9:PW01} is
promising, it is currently limited to $A\leq 10$ and not
applicable to continuum states. Cluster models have been often
used for light systems so far \citep[see][and references therein]{sec9:DR06,sec9:DesBook}.
They assume that the nucleons are grouped
in clusters and use cluster wave functions defined in the shell
model and computed with an adapted effective nucleon-nucleon
force. The Resonating Group model (RGM) and the Generator
Coordinate Method (CGM) are two equivalent implementations
differing in the definition of the relative wave function of the
clusters \citep{sec9:DR06,sec9:DesBook}.

\subsection{Statistical Model}
\label{sec:9:highdensitymodels}

\begin{figure}[t] 
\includegraphics[width=\textwidth]{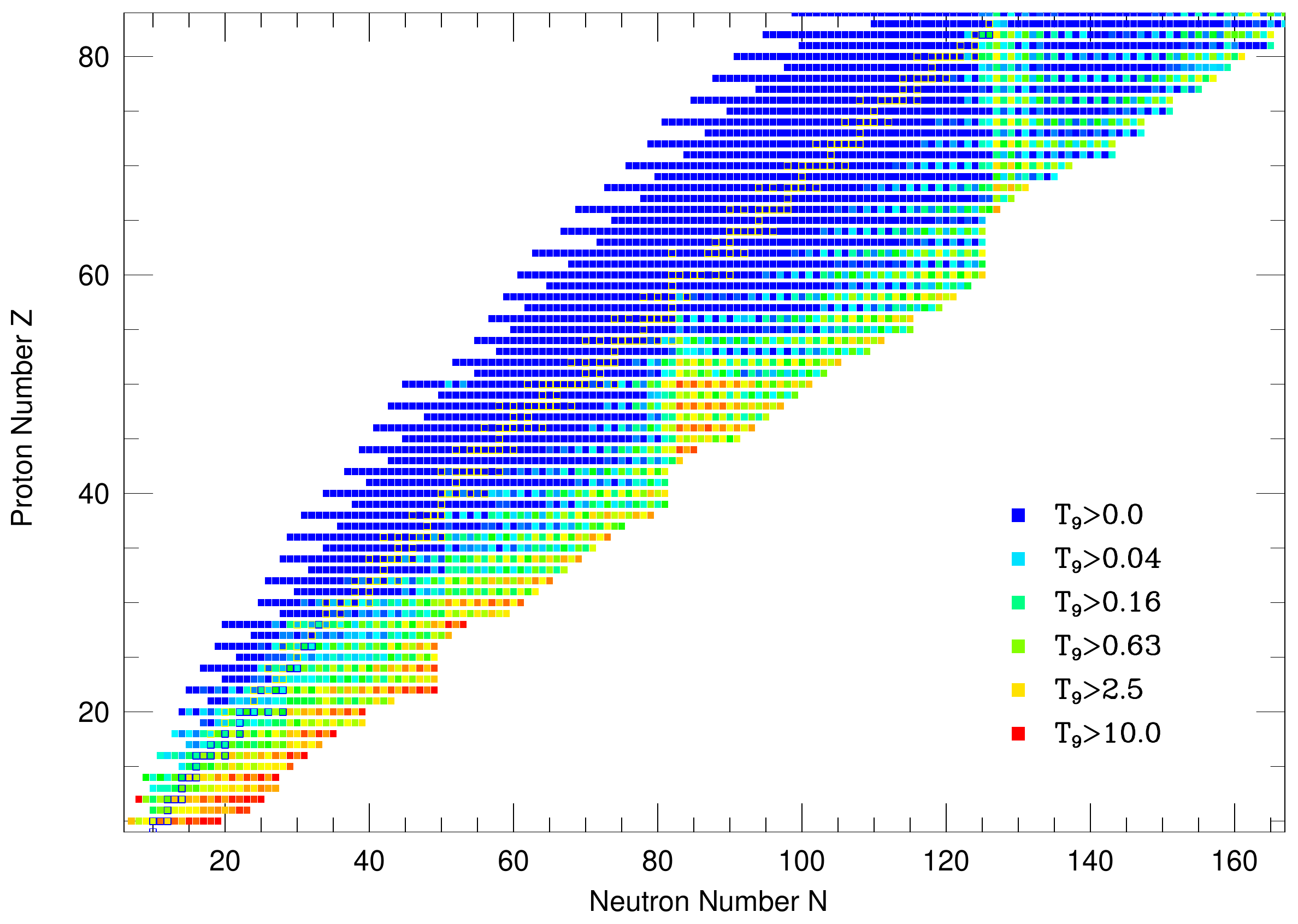}
\caption{Applicability of the Hauser-Feshbach model to calculate astrophysical reaction rates for neutron-induced reactions: Shown are the stellar temperatures above which the rate can be calculated from Hauser-Feshbach cross sections \citep[reprinted from][with permission]{sec9:RTK97}.
}
\label{Fig:HF}
\end{figure}

In systems with high level density $\rho(J,\pi,E)$, individual
resonances cannot be resolved anymore and an average over the
overlapping resonances can be used instead (Fig.~\ref{Fig:HF}). Further assuming that
the relative phases are randomly distributed, interferences will
cancel and a simple sum of Breit-Wigner contributions can be
replaced by a level-density weighted sum of averaged widths
$\left< \Gamma \right>$ over all spins $J$ and parities $\pi$
\citep{sec9:DR06,sec9:GH92}
\begin{equation}
\sigma ^x (E) \propto  \frac{1}{(2J_x+1)(2J_\mathrm{proj}+1)} \nonumber \end{equation} \begin{equation}
  \times \sum _{J,\pi } \left[ (2J+1) \rho(J,\pi,E_\mathrm{c}) \right. \nonumber \end{equation} \begin{equation}
\times \left< \Gamma^x_\mathrm{pro} (\{J_x,\pi_x\}\rightarrow \{J,\pi\},E) \right> \nonumber \end{equation} \begin{equation}
\frac{\left< \Gamma_\mathrm{b} \left(\sum_{J_\mathrm{fin},\pi_\mathrm{fin},E_\mathrm{fin}} \left( \{J,\pi\} \rightarrow \{J_\mathrm{fin},\pi_\mathrm{fin}\},E_\mathrm{fin}\right) \right) \right>}{\left< \Gamma_\mathrm{tot} \right>} \nonumber \end{equation} \begin{equation}
\left. \times W(J,\pi,E_\mathrm{c}) \right] \quad, \end{equation} \begin{equation}
E_\mathrm{c}=E+E_\mathrm{sep,pro}-E_x \quad,\end{equation} \begin{equation}
E_\mathrm{fin}=E_\mathrm{c}-E_\mathrm{sep,fin}-E_{x,\mathrm{fin}} \quad.
\end{equation}

This is called the \textit{Hauser-Feshbach} or \textit{statistical
model of compound reactions} \citep{sec9:HF52}. Width fluctuation
corrections $W$ account for non-statistical correlations but are
only important close to channel openings \citep{sec9:Eric60}. The
separation energy $E_\mathrm{sep,pro}$ of the projectile in the
compound system determines at which energy $E_\mathrm{c}$ the
compound system is formed. The averaged width of the exit channel
$\left< \Gamma_\mathrm{b} \right>$ usually includes a sum over
energetically possible final states at energy $E_{x,\mathrm{fin}}$
or an integral over a level density of the final system when
individual states are not known or numerous. For capture, compound
and final system are identical. The averaged widths are related to
transmission coefficients $\mathcal{T}=2\pi \rho \left< \Gamma
\right>$. The latter are calculated from the solution of a
(radial) Schr\"odinger equation using an optical potential. (It is
to be noted that these potentials differ from the ones employed
for low-density systems described in Sec.\
\ref{sec:9:lowdensitymodels}.)

The challenge for nuclear astrophysics lies in the determination
of globally applicable descriptions of low-energy optical
potentials as well as level densities, masses (determining the
separation energies), and spectroscopy (energies, spins, parities)
of low-lying excited states, to be applied for a large number of
nuclei at and far from stability. For details on the different
properties and the remaining open problems in their treatment
\citep[see, e.g.,][ and references therein]{sec9:DR06,sec9:RTK97,sec9:fiss2,sec9:thalys,raureview}.
For a general discussion of the applicability
of the statistical model, see \citet{sec9:RTK97,sec9:RT00,raureview}.

\section{Experimental Facilities and Techniques}
\label{sec:9:expfac}

The experimental determination or verification of nuclear reaction
rates requires a large variety of facilities and techniques. This
is in particular true if one wants to establish experimentally
reaction rates associated with the production of long-lived
radioactive isotopes associated with galactic gamma sources.
Nuclear astrophysics related experiments include low energy high
intensity accelerator measurements with stable beams to study
charged particle reactions of relevance for quiescent stellar
burning which may possibly lead to the production of $^7$Be,
$^{22}$Na and $^{26}$Al. High flux neutron beam studies to explore
neutron induced reactions for the weak and main s-process which
can be associated with the production of long-lived radioactive
isotopes such as $^{60}$Fe and $^{98}$Tc. Real and virtual photon
beams are increasingly used for probing nuclear reactions
associated with explosive nucleosynthesis events such as the
p-process but can also be used to probe indirectly neutron capture
reactions associated with the s-process. Intense radioactive beams
are the primary tools for exploring nuclear reactions and decay
mechanisms far of stability which are expected to occur in
explosive stellar environments and may lead to the production of
long-lived radioactive elements such as $^{18}$F, $^{26}$Al,
$^{44}$Ti and $^{56}$Ni.

\subsection{Low-energy Facilities, Underground Techniques}
\label{sec:9:explowunder}

Low energy charged particle measurements belong to the most
challenging experiments in nuclear astrophysics. The cross
sections need to be measured at the extremely low energies
associated with the Gamow range of quiescent stellar burning. This
requires to determine the cross sections of proton capture
reactions for hydrogen burning in main sequence stars at energies
well below 100 keV. Measurements for helium burning in red giant
stars need to be explored in the 200 keV to 500 keV range and
heavy ion fusion reactions in subsequent stellar evolution phases
need to be measured near 1 MeV to 2 MeV center of mass energy. The
cross sections are extremely low, typically in the femto-barn
range, which requires a long time, in excess of days, to
accumulate a statistical relevant amount of reaction yield data.
Typical experimental techniques are summarized in the text book
literature \citep{sec9:Il07} and will not be discussed here.

The critical issue with low cross sections is that the yield of
reaction events is low in the detectors measuring the
characteristic gamma or particle radiation produced. This requires
using high efficiency detector material with high resolution to
separate the characteristic events from random background events.
High beam intensity is desired to increase the event rate, however
it may also increase beam induced background on target impurities
and is limited by target stability.

The second critical issue is the background rate in the detector.
There are typically three different kind of background, cosmic ray
induced background in the detector environment, natural long-lived
radioactivity in the detector material and the surrounding
environment, and beam induced background on low Z target
impurities and beam defining slits or apertures. This background
must be reduced as far as possible to identify reaction events in
the spectrum.

Cosmic ray induced background affects the spectra up to very high
energies and makes it difficult to extract weak signals. That
background is the most important to remove. Natural environmental
background will be strong in an underground environment except for
salt mine locations. But the characteristic $\gamma$ lines are
mainly below 3 MeV and can be shielded locally. Neutron background
is more difficult to absorb and needs special shielding
arrangements. Beam induced background depends critically on the
target as well as the choice and preparation of the target
material. It is difficult to suppress and may require active
shielding procedures.

This can be done by identifying the event electronically by its
particular characteristics such as coincidence conditions in a
particular decay sequence, pulse shape or timing conditions and
reject the background events which do not fulfill these
requirements. This can lead to active background suppression by up
to three orders of magnitude \citep{sec9:RCA05,sec9:CBC08}. While
this clearly helps in many cases a more efficient background
reduction is desired.

\index{LUNA facility} \index{TRIUMF facility} \index{DUSEL laboratory}
The high energy cosmic ray induced background can be most
successfully suppressed by operating the experiments in a deep
underground environment where the cosmic ray flux is heavily
reduced. This was demonstrated with the installation of the LUNA
accelerator facility at the Gran Sasso deep underground laboratory
in Italy. The cosmic ray induced background was successfully
removed and several critical reactions of the pp-chains and the
CNO cycles were successfully measures in the or near the Gamow
energy range \citep{sec9:Co09}. As a consequence of this
successful operation new underground accelerator facilities are
being proposed or planned which would allow to cover reactions
over a wider energy range than available at LUNA. This is of
particular importance for an improved R-matrix analysis and
extrapolation. Higher energies are also of great relevance for the
underground measurements of $\alpha$ capture reactions and stellar
neutron sources in helium burning. In particular it will also
improve the chances for pursuing heavy ion fusion reaction studies
towards lower energies. There are presently three major
initiatives for the construction of new underground accelerator
facilities. The proposal to establish an underground accelerator
facility ELENA at the Boulby salt mine in the UK seeks to take
advantage of the reduced level of neutron and natural activity in
a salt environment. The disadvantage will be the reduced depth
level compared to the Gran Sasso location. The second proposal is
for the development of a two accelerator facility DIANA at the
DUSEL underground laboratory at Homestake mine in South Dakota.
The third proposal in debate is the construction of an accelerator
facility in an abandoned train tunnel in the Pyrenees mountains at
Canfranc, Spain. With these facilities the community hopes to
address the new and critical questions about stellar reaction
cross sections and provide the final answer on the nuclear engine
of stellar evolution.

However, it has been demonstrated that alternative inverse
kinematics methods are a very powerful tool in reducing the
background. They are based on the technique of using a high
intensity heavy ion beam on a hydrogen or helium gas target and
separate the heavy ion recoil reaction products from the primary
beam through a high resolution electromagnetic mass separator
system from the primary beam. This method has been demonstrated to
be successful at a number of different separator facilities such
as DRAGON at TRIUMF, Vancouver \citep{sec9:Voc07} (Fig.~9.2) and ERNA at the
Ruhr University Bochum \citep{sec9:DiL09}.

\begin{figure}[t] 
\includegraphics[width=\textwidth]{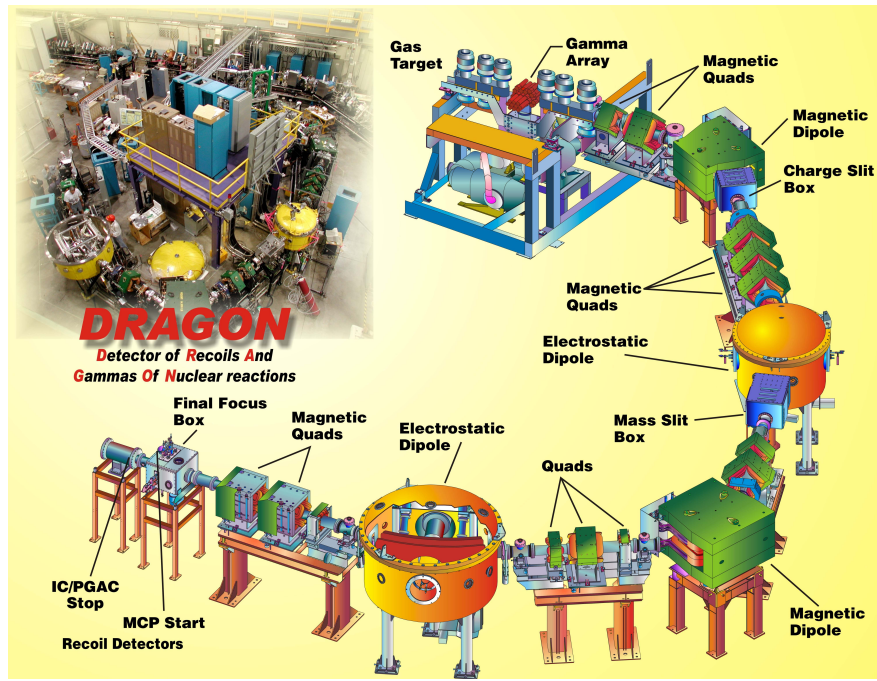}
\caption{The \emph{Dragon} facility in Vancouver, Canada, is an example of nuclear experiment facilities, now aimed at experiments for astrophysically-relevant reactions: Radioactive isotopes can be selected and accelerated to form projectiles for such reactions of interest.
}
\label{Fig:Dragon}
\end{figure}  

The detection of the recoiling charged particle has a clear
efficiency advantage compare to the gamma detection. The possible
detection of the gamma rays in coincidence with the reaction
products reduces dramatically the backgrounds. However, there are
several experimental challenges associated with using recoil
separators. At the low stellar energies, the energy spread and the
angular aperture are much larger than the acceptance of any of the
cited existing recoil separators. In order to measure an absolute
cross section the transmission of the recoils should ideally be
100\% or exactly known. It is also necessary to know precisely the
charge state distribution of the recoil products. In addition,
since the primary beam intensity is typically many orders of
magnitude larger than the recoiling reaction products, a large
spatial separation between the reaction products and the beam is
required, which is difficult to realize for beams with a large
energy spread. Therefore, solar fusion reactions are particularly
challenging to measure with recoil separators and are typically
used for higher energies and for the helium or heavy ion burning
reactions.

Dedicated next generation separators for low energy nuclear
astrophysics studies with stable ion beams coming on line are the
ST.GEORGE facility in Notre Dame \citep{sec9:Cou08} and the
modified and upgraded ERNA facility at CIRCE in Caserta, Italy.
Both separators feature large angular and energy acceptance and
will be equipped with high density gas jet targets, which will
ensure a well defined interaction region.

\subsection{Laboratory Neutron Sources}
\label{sec:9:nexpneutrons}

Many of the observed or anticipated long lived radioactive
isotopes in our galaxy are produced by neutron induced
nucleosynthesis in the weak or main s-process taking place in
helium and carbon burning stellar environments. This includes
$^{41}$Ca, $^{60}$Fe, $^{63}$Ni, but also more massive isotopes
such as $^{98}$Tc and $^{99}$Tc and possibly numerous long lived
isomers.
\index{isotopes!98Tc} \index{isotopes!41Ca} \index{isotopes!60Fe} \index{isotopes!63Ni}

The study of neutron induced stellar reactions leading to the
production of such isotopes requires high intensity neutron
sources with a well defined energy distribution to determine the
reaction cross sections at stellar energies of a few keV. Neutrons
in that energy range can be produced in several ways. Nuclear
reactions such as $^7$Li(p,n) or $^3$H(p,n) with high intensity
proton beams provided by low-energy particle accelerators offer
the possibility of tailoring the neutron spectrum to the energy
range of interest; this has the advantage of low backgrounds. A
particularly successful approach is to simulate a quasi-stellar
neutron spectra in the laboratory. In bombarding thick metallic
lithium targets with protons of 1912 keV, the resulting neutrons
exhibit a continuous energy distribution with a high-energy cutoff
at E$_n$ = 106 keV and a maximum emission angle of 60 degrees. The
angle-integrated spectrum corresponds closely to a
Maxwell-Boltzmann distribution for kT = 25 keV \citep{sec9:Rat88}.
Hence, the reaction rate measured in such a spectrum yields
immediately the proper stellar cross section.

Higher intensities can be achieved via photon production by
bombarding heavy-metal targets with typically 50-MeV electron
beams from linear accelerators. When these energetic neutrons are
slowed down in a moderator, the resulting spectrum contains all
energies from thermal energy to nearly the initial energy of the
electron beam. Since the astrophysical relevant range corresponds
to only a narrow window in this spectrum, background conditions
are more complicated and measurements need to be carried out at
larger neutron flight paths. In turn, the longer flight paths are
advantageous for neutron-resonance spectroscopy with high
resolution.

The most intense keV neutron flux is produced by spallation
reactions. The LANSCE facility at Los Alamos is particularly
suited for neutron TOF work due to the favorable repetition rate
of only 12 Hz \citep{sec9:Lansce}, and because the accumulation of
a number of beam pulses in an external storage ring yields
extremely intense neutron bursts. Accordingly, excellent
signal-to-background ratios can be achieved. The n\_ToF facility at
CERN provides high intensity neutron beam pulses with a lower
repetition rate of 0.4 Hz \citep{sec9:ntof}. This has proved
highly advantageous for a large number of experimental neutron
capture studies along the s-process path.
\index{LANSCE facility} \index{CERN}

The experimental methods for measuring (n,$\gamma$) cross sections
fall into two groups: TOF techniques based on the detection of the
prompt capture $\gamma$ rays and activation methods.

The TOF techniques can be applied in measurements of most stable
nuclei but require a pulsed neutron source to determine the
neutron energy via the flight time between target and detector.
Capture events in the samples are identified by the prompt
$\gamma$ ray cascade in the product nucleus.

The best signature for the identification of neutron capture
events is the total energy of the capture gamma cascade, which
corresponds to the binding energy of the captured neutron. Hence,
accurate measurements of (n,$\gamma$) cross sections require a
detector that operates as a calorimeter with good energy
resolution and is insensitive to neutron exposure. In the gamma
spectrum of such a detector, all capture events would fall in a
line at the neutron binding energy (typically between 5 and 10
MeV), well separated from the gamma-ray backgrounds that are
inevitable in neutron experiments. Such detectors have been
successfully developed at the various laboratories using arrays of
4$\pi$ BaF$_2$ scintillator detectors with a large number of
independent detector modules \citep{sec9:Hei01} and have emerged
as standard technology for these kinds of measurements.

A completely different approach to determining stellar
(n,$\gamma$) rates is activation in a quasi-stellar neutron
spectrum. Compared with the detection of prompt capture gamma
rays, this method offers superior sensitivity, which means that
much smaller samples can be investigated. Since it is also
selective with respect to various reaction products, samples of
natural composition can be studied instead of the expensive
enriched samples required by the TOF techniques. However, the
activation technique is restricted to cases where neutron capture
produces an unstable nucleus, and it yields the stellar rate only
for two thermal energies at kT = 25 and 52 keV. This method is
however particulary powerful in obtaining cross sections for
reactions producing long-lived radioactive materials which can be
identified by their particular decay characteristics and
signature. This activation technique has been used for a variety
of measurements. The technique can be applied to short-lived
products with half-lives in the millisecond range and allows for
cross section measurements with uncertainties of a few percent.

\subsection{Accelerator Mass Spectroscopy}
\label{sec:9:expams}


Classical activation techniques require a characteristic decay
signal associated with the decay pattern or the half-life of the
produced radioactive isotope. This can be difficult in cases where
no characteristic gamma or particle decay pattern exists or where
the decay analysis of the $\beta$ decay signal is prohibited by
high background activity. In these cases activation analysis
through accelerator mass spectrometry (AMS) offers a powerful tool
\index{accelerator mass spectroscopy}
to measure cross sections through ultra-low isotope-ratio
determination. The AMS method was successfully introduced for the
study of the neutron-capture cross section of
$^{62}$Ni(n,$\gamma$)$^{63}$Ni \citep{sec9:Nas05}, and extended to
other neutron and charge-particle-induced reactions, such as
$^{25}$Mg(p,$\gamma$)$^{26}$Al \citep{sec9:Ara06} and
$^{40}$Ca($\alpha$,$\gamma$)$^{44}$Ti \citep{sec9:Nas06}. \index{isotopes!44Ti} \index{isotopes!26Al}

In these cases samples were either irradiated in a neutron
spectrum resembling a stellar Maxwell Boltzmann distribution or by
charged particles of well known energies. After the irradiation
the samples must be chemically treated to extract the radioactive
reaction products. This requires some time and limits AMS
activation studies to more longer lived isotopes. Since isotopic
and isobaric interferences may represent a major challenge in AMS
measurements of irradiated samples, extensive background studies
for these isotopes are always necessary prior to the irradiations
in order to demonstrate that the required sensitivity can be
reached. In AMS, negative ions are extracted from an ion source
which have to pass a low energy mass spectrometer prior to
entering a tandem accelerator. When passing the stripper, positive
ions are produced while within this stripping process molecular
isobars are destroyed. One positive charge state is selected with
a second (high-energy) mass spectrometer system which is optimized
for mass, charge and isobar separation trough possible combination
of dipole magnet separators, Wien-filters, and more recently
magnetic gas filled separators for improved isobar separation.
With such a system the concentration ratio of the radioisotope is
determined relative to a stable isotope by measuring the number of
radionuclides relative to the current of the isotopic ions in
front of the detector, after adjusting the injector magnet,
terminal and Wien-filter voltage appropriately. By measuring
relative to a standard sample of known isotopic ratio, factors
like stripping yields and transmissions mostly cancel.

The difficulties with AMS experiments is in the chemical
preparation of the sample and the sufficient separation of the
extracted radioactive ions from background events. While AMS is a
widely established method with many applications, the analysis of
the very limited number of radioactive products from low cross
section reactions remains challenging. Systematic studies are
necessary to reduce possible uncertainties.

Dedicated AMS facilities with an established nuclear astrophysics
program are the Vienna Environmental Research Accelerator (VERA)
\citep{sec9:Stei02}, the Center for Isotopic Research on Cultural
and Environmental Heritage (CIRCE) in Caserta/Italy
\citep{sec9:Ter07} or the Munich Tandem accelerator facility
\citep{sec9:Kni00} which is optimized for the analysis of more
massive radioactive isotopes. A new AMS program is presently being
developed utilizing the Notre Dame tandem accelerator
\citep{sec9:Rob08}.

\subsection{Radioactive Beam Techniques}
\label{sec:9:exprib}


The development of radioactive accelerated beams for low energy
nuclear astrophysics experiments has been one of the large
challenges of the field. The experimental study of nuclear
reactions and decay processes far of stability is necessary for
the understanding of explosive nucleosynthesis processes such as
the rp-process in cataclysmic binary systems or the r-process and
p-process in the supernova shock front. These processes can in
particular contribute to the production of long-lived galactic
radioactivity by primary reaction or also by secondary decay
processes from the reaction path towards the line of stability.
\index{radioactivity!beam experiments}

For the purpose of studying the origin of long-lived radioactive
isotopes in astrophysical environments radioactive beams are
utilized in two ways, for producing long-lived targets by
implantation for subsequent irradiation with neutron, charged
particle or possibly intense photon beams or for direct reaction
measurement in inverse kinematics on light ion target materials.
The later approach requires well defined mono-energetic and
intense radioactive beams and a detection system for light or
heavy recoil reaction products.

The main challenge in this approach is to produce a sufficiently
high intensity of radioactive beams which have to be produced
on-line as a secondary reaction product. This requires high cross
sections for the production process and high primary beam
intensities. A variety of different approaches has been chosen in
the past to optimize the production efficiency and maximize the
intensity of the radioactive beams. A technique developed for
small scale facilities is the selection of specific nuclear
reactions tailored for the on-line production of radioactive beams
at optimum conditions. The secondary particles can be used for
subsequent nuclear reaction studies after blocking and separation
from the primary beams \citep{sec9:twinsol}. The efficient
separation of a suitably high intensity beam of radioactive
species is the most challenging problem for this approach.

An alternative approach is the ISOL (Isotope Separation On-Line)
\index{isotopes!separator ISOL}
technique where high energy protons are used to bombard heavy ion
targets for producing a large number of radioactive species
through spallation processes. These isotopes diffuse out of the
target into an ion source for being charged and re-accelerated for
secondary beam decay or reaction experiments. The method has been
proven to be very powerful over the years but is limited to
isotopes with lifetimes appreciably longer that the time necessary
for the diffusion transport and ionization process. This can be
different for different elements because of the associated
chemical processes between the isotopes and the surrounding
environment.

The third approach is based on the use of energetic heavy
projectiles bombarding light or target nuclei fragmenting on
impact. This fragmentation process generates a cocktail beam of
many radioactive species which move forward with high velocity
because the initial momentum of the primary particles is
maintained. For experiments with a specific secondary particle, it
must be selected by fragment separator systems which separate and
focus the isotopes by magnetic fields and energy loss
characteristics in heavy wedge materials. For nuclear astrophysics
related experiments the fast beam particles need to be slowed down
by energy loss in gas or solid material and re-accelerated to
energies corresponding to the temperatures in the explosive
stellar scenarios.

There is a number of laboratories which have focused on nuclear
reaction studies with radioactive beams. The first fully operating
radioactive beam laboratory based on the ISOL principle was the
coupled cyclotron facility at Louvain la Neuve which did a number
of successful radioactive beam studies of relevance for
investigating the the production of $^{18}$F in novae \index{isotopes!18F}
\citep{sec9:103}. These measurements were complemented by
measurements at the HRIBF facility at Oak Ridge using intense
\index{HRIBF facility}
$^{18}$F beams \citep{sec9:102}. Both facilities produce the
radioactive species by nuclear reactions on thin production
targets, with the reaction products being transported into an ion
source for producing and subsequently accelerating the secondary
beam. The intensity is largely limited by target technology and
beam transport and re-ionization efficiency.

The premier ISOL radioactive beam facility is ISAC at TRIUMF
Canada. The primary 600 MeV proton beam is provided by the TRIUMF
cyclotron. The reaction products are post-accelerated in an RFQ SC
LINAC accelerator combination to energies of 0.3 to 3 MeV/u. ISAC
has successfully performed a number of radioactive beam
experiments of relevance for explosive hydrogen and helium
burning. Most notable a direct study of \index{isotopes!22Na}
$^{21}$Na(p,$\gamma$)$^{22}$Mg in inverse kinematics to probe the
production mechanism of $^{22}$Na in Ne nova explosion
environments \citep{sec9:21Na}. The facility also runs a
successful program with stable beams which was utilized to
investigate the production of $^{44}$Ti \citep{sec9:Voc07}. \index{isotopes!44Ti}
Presently a number of studies associated with the production of
the long-lived $\gamma$ emitter $^{26}$Al are being performed. \index{isotopes!26Al}

Other ISOL based radioactive beam facilities such as Spiral
facility at GANIL in  Caen, France or REX-ISOLDE at CERN have been
used to perform interesting experiments for nuclear astrophysics
but have been less concerned with the question of nuclear
production mechanisms for long lived cosmic gamma emitters.

There have been a number of fast radioactive beam facilities with
scientific programs in nuclear astrophysics primarily aimed at the
study of nucleosynthesis processes far off stability. However the
rapid new developments in fast beam physics promises a number of
new experimental opportunities which can provide benefits for
studying reactions associated with the production of long-lived
gamma emitters in explosive nucleosynthesis events.

There are currently four major fragmentation facilities in the
world: GANIL and GSI in Europe, NSCL/MSU in the US and RIKEN in
\index{GANIL facility} \index{GSI facility} \index{RIKEN facility}
Japan. They are all based on Heavy Ion accelerators which operate
in complementary energy domains. Because of the high energy of the
fragment products low energy reaction experiments for nuclear
astrophysics are not possible but the development of indirect
techniques to determine critical reaction or decay parameters has
been the primary goal. In the context of long-lived isotopes of
astrophysical interest a major contribution was the development of
fast beams such as $^8$B at NSCL/MSU, RIKEN, and GSI for utilizing
Coulomb dissociation techniques for probing critical reactions
such as $^7$Be(p,$\gamma$)$^8$B. The NSCL and RIKEN also
successfully developed a $^{44}$Ti beam for new measurements of
its half-life \citep{sec9:Goe98}. More half-life measurements of
long-lived isotopes such as $^{60}$Fe are presently underway to
re-evaluate these critical parameters.

\section{Specific Experiments}
\label{sec:9;expexamples}

The complexities of the experiments and the uncertainties in the
experimental results affect the reliability of model predictions
on the nucleosynthesis of long-lived radioactive species. In
particular recent studies of critical nuclear reactions and decay
processes exhibit considerable differences to earlier studies
which so far have been the reference point for nucleosynthesis
simulations and predictions for long-lived radioactive isotope
abundances in stellar burning environments. It is therefore
important to carefully evaluate the experimental results and
clarify possible discrepancies and inconsistencies in the data.
This section will discuss the present status of the experimental
reaction rates and evaluate future opportunities to improve the
existing data base.

\subsection{Experiments with Stable Beams}
\label{sec:9:expexstable}

Many of the long-lived radioactive gamma emitters in our universe
have been produced by radiative capture reactions on stable
isotopes. The best known examples are $^{26}$Al, which is
primarily formed by proton capture on stable $^{25}$Mg isotopes,
$^{25}$Mg(p,$\gamma$)$^{26}$Al, and $^{44}$Ti which is most likely \index{isotopes!44Ti}
produced via alpha capture on stable $^{40}$Ca isotopes,
$^{40}$Ca($\alpha$,$\gamma$)$^{44}$Ti. Extensive measurements
using in-beam $\gamma$ spectroscopy techniques have been made for
both reactions and have formed the basis for earlier reaction rate
compilations.

The low energy reaction cross section of
$^{25}$Mg(p,$\gamma$)$^{26}$Al is characterized by several \index{isotopes!26Al}
resonances with energies between 30 keV and 400 keV . The reaction
rate is directly correlated to the strengths $\omega\gamma$ of the
resonances. The strengths for the resonances above 190 keV have
been determined from the on-resonance thick target yield in
radiative capture measurements \citep{sec9:eli79,sec9:ili90}. The
strengths of lower energy resonances are estimated on the basis of
single particle transfer reaction studies. Of particular
importance are three resonances at 90 keV, 130 keV and at 190 keV
which determine the reaction rate at temperatures typical for
stellar hydrogen burning in AGB stars  \index{stars!AGB} and nova explosions. Because
the low energy radiative capture measurements have been
handicapped by cosmic ray induced background, an alternative
measurement was done using the AMS technique to analyze the number
of $^{26}$Al reaction products after irradiation at resonance
energies \citep{sec9:Ara06}. The experiment was successful and
confirmed the resonance strengths of the known resonances at 304
keV, 347 keV, and 418 keV resonance energy. However the results
indicated a substantially lower strength for the critical
resonance at 190 keV. This would reduce the reaction rate by about
a factor of five at the temperature range between 0.2 and 1.0 GK.
This result introduced a large uncertainty in the reaction rate
which motivated a new experimental study at LUNA in the Gran Sasso
laboratory using in-beam gamma spectroscopy techniques with a
variety of high efficiency and high resolution gamma detector
devices. The measurements confirmed earlier gamma spectroscopy
studies of the strengths of higher energy resonances
\citep{sec9:eli79,sec9:ili90} tabulated in the NACRE compilation
\citep{sec9:nacre}. The new results are being prepared for
publication. Parallel to the gamma spectroscopy measurement, the
irradiated samples were analyzed for their $^{26}$Al content using
AMS techniques. The AMS measurements were performed at the CIRCE
facilities. Excellent agreement is demonstrated for the resonance
at 304 keV, additional experiments are being pursued for lower
energy resonances to address the inconsistencies in the strength
determination for the 190 keV resonance.

The $^{40}$Ca($\alpha,\gamma$)$^{44}$Ti reaction is considered to
be one of the major production reactions for $^{44}$Ti in
supernova shock front nucleosynthesis. The cross section for this
radiative capture process has been explored in a number of in-beam
gamma spectroscopy studies down to center of mass energies of 2.5
MeV \citep{sec9:di71,sec9:coo77}. The cross section is
characterized by a large number of resonances and the initial
reaction rate determinations were based on an analysis of
resonance strengths. Despite the high level density in $^{44}$Ti,
it was noted that the experimental reaction rate is substantially
smaller than the reaction rate based on statistical model Hauser
Feshbach predictions \citep{sec9:Rau00}. The reaction was studied
independently using a thick He-gas cell target and counting the
long lived $^{44}$Ti reaction products by AMS techniques
\citep{sec9:Nas05} to determine the integral yield over an energy
range of 1.7 to 4.2 MeV. The extracted reaction rate is
substantially higher than the ones discussed in the literature
\citep{sec9:Rau00}. A more recent study of the reaction using
inverse kinematics techniques was performed at the ISAC facility
at TRIUMF, Vancouver, separating the $^{44}$Ti reaction products
on-line with the DRAGON recoil separator. The measurements covered
the energy range of 2.3 MeV to 4.2 MeV (center of mass) in more
than 100 small energy steps. The extracted yield was mostly
interpreted as on-resonance resonance thick target yield and
translated to a resonance strength. There are large uncertainties
associated with this approach, in particular with the
determination of the resonance energies, which have not been
unequivocally determined in the experiment. In some cases several
of the quoted resonances agree with previously identified states,
in other cases it needs to be confirmed that the observed yields
really correspond to additional resonances and do not originate
from tail contributions of resonant yield curves associated with
the different states. As far as the resonance levels which have
been observed in both studies are concerned the published
strengths are comparable to each other. Nevertheless the reaction
rate suggested by Vockenhuber et al. \citep{sec9:Voc07} is larger
by more than a factor of two than the rates projected on the basis
of the in-beam gamma spectroscopy measurements, but it is in
agreement with the projections by Rauscher et al.
\citep{sec9:Rau00}. The difference is mainly due to the difference
in resonance numbers. While the resonance identification in
previous work was based on a careful analysis of the particular
gamma decay characteristics of the observed levels, the analysis
of the recoil data is insufficient in providing information to
differentiate between different resonances. It cannot be excluded
that the number of identified states are overestimated; a more
detailed gamma spectroscopy study with thin targets is therefore
highly advisable to remove the existing uncertainties.

\subsection{Experiments with Neutron Beams}
\label{sec:9:expexneutrons}

A particularly interesting case is the origin of the long-lived
gamma emitter $^{60}$Fe. Its characteristic $\gamma$-radioactivity
has been observed with the INTEGRAL gamma ray telescope in
supernova remnants near the solar system. These observations are
complemented by recent AMS studies which suggest high $^{60}$Fe
abundance in deep sea ferromanganese sediments \citep{sec9:Kni04}.
These $^{60}$Fe observations have been interpreted as indication
for the existence of a recent ($\approx$ 3 million years) supernova
event in the solar system vicinity. A more quantitative
interpretation of the time and distance of the proposed supernova
event requires a detailed knowledge of the nucleosynthesis history
of $^{60}$Fe.

The radioactive $^{60}$Fe isotope is produced by a sequence of \index{isotopes!60Fe}
neutron capture reactions of stable iron isotopes such as
$^{58}$Fe(n,$\gamma$)$^{59}$Fe(n,$\gamma$)$^{60}$Fe, the
production rate and final abundance of the long-lived $^{60}$Fe
depends on the reaction rate of these feeding processes as well as
on the rate of the $^{60}$Fe(n,$\gamma$)$^{61}$ depletion
reaction. No experimental information are available about the
associated cross sections except for the neutron capture reaction
$^{58}$Fe(n,$\gamma$)$^{59}$Fe. Present simulations of the
$^{60}$Fe nucleosynthesis rely entirely of statistical model
predictions of the neutron capture rates. Because of the
relatively low level density in the associated $^{60}$Fe,
$^{61}$Fe compound nuclei these model predictions are unreliable
and need to be tested experimentally. This is underlined by the
direct comparison between the experimental cross sections for
neutron capture on the stable isotopes $^{56}$Fe, $^{57}$Fe, and
$^{58}$Fe which were all measured through neutron activation
techniques and theoretical Hauser Feshbach predictions which show
considerable discrepancies in particular in the cases of
$^{56}$Fe(n,$\gamma$)$^{57}$Fe and $^{57}$Fe(n,$\gamma$)$^{58}$Fe.
For $^{58}$Fe(n,$\gamma$)$^{59}$Fe on the other hand , the
agreement seems reasonable well but that cannot be extrapolated
towards neutron captures on the more neutron rich Fe isotopes
which are subject of the here proposed measurements.

Particularly important is the determination of the reaction rate
of $^{59}$Fe(n,$\gamma$)$^{60}$Fe since it competes directly with
the $^{59}$Fe $\beta$-decay which would by-pass the production of
$^{60}$Fe. A direct measurement of this critical reaction in the
traditional activation or time of flight spectroscopy technique is
not feasible because the target is radioactive and only small
amounts can be accumulated. these small amounts nevertheless
produce a large background activity level, which would prohibit
any of the described methods. The cross section for the ground
state decay branch of $^{59}$Fe(n,$\gamma_0$)$^{60}$Fe can however
be investigated using inverse $^{60}$Fe($\gamma$,n)$^{59}$Fe
Coulomb dissociation techniques. The $^{60}$Fe beam can be
produced by fragmentation of a heavy ion such as $^{64}$Ni on a
light Be target at an energy of 500 MeV/u. The $^{59}$Fe recoil
products, and the released reaction neutrons, as well as $\gamma$
rays can be detected with reasonable 200 keV resolution using a
combination of a magnetic separator system and a neutron detector
wall. This allows particle identification of all reaction
products.

\subsection{Experiments with Radioactive Beams or Targets}
\label{sec:9:expexrib}

The depletion processes of long-lived radioactive isotopes
includes the natural decay. Simulating this branch requires not
only a good knowledge of the laboratory lifetime but also of the
nature of the decay process since extreme environmental effects
can change the decay rates drastically. In terms of $\beta$ decay,
the decay can be accelerated through the decay of thermally
excited states as in the case of $^{26}$Al. For decay through
electron capture, the decay can be slowed down since the nuclei
are completely ionized and the electrons have to be captured from
the stellar plasma rather than from the inner K- or L-shell of the
atom. This affects in particular the lifetime of $^{44}$Ti, which \index{isotopes!44Ti}
primarily decays by electron capture.

Often the depletion is primarily driven by nuclear reactions, such
as $^{22}$Na(p,$\gamma$)$^{23}$Mg, $^{26}$Al(p,$\gamma$)$^{27}$Si,
$^{44}$Ti($\alpha$,p)$^{47}$V, or $^{60}$Fe(n,$\gamma$)$^{61}$Fe,
but also capture reactions on shorter-lived excited configurations
of these nuclei are possible, such as
$^{26}$Al$^*$(p,$\gamma$)$^{27}$Si. There are two possibilities
for experimental studies of the reaction cross sections. The first
one is based on the production of highly enriched long-lived
radioactive targets, which can be prepared through standard
chemical target preparation techniques using externally bred
radioactive material, or by implantation of radioactive ions at
low energy ISOL facilities. The disadvantage of both techniques is
that the actual $\gamma$ measurements have to be performed in a
high radiation background environment produced by the sample
itself.

Nevertheless, earlier measurements of reactions such as
$^{22}$Na(p,$\gamma$)$^{23}$Mg \citep{sec9:Seu89} and \index{isotopes!22Na}
$^{26}$Al(p,$\gamma$)$^{27}$Si \citep{sec9:Buc81} relied entirely
on this approach. In both cases a large number of resonances were
detected and the resonance strength determined for calculating the
reaction rates. The results for $^{22}$Na(p,$\gamma$)$^{23}$Mg
were confirmed by new direct measurements using improved target
and detection techniques \citep{sec9:Ste96}, resulting in the
observation of an additional low energy resonance at lower
energies. Complementary spectroscopy techniques such as the study
of the $\beta$-delayed proton decay of $^{23}$Al
\citep{sec9:Per00} and the heavy ion reaction induced $\gamma$
decay of proton unbound states in $^{23}$Mg \citep{sec9:Jen04}
provided additional nuclear structure information which led to the
reduction of uncertainties in the reaction rate.

The situation is similar with $^{26}$Al(p,$\gamma$)$^{27}$Si;
after the initial study with radioactive targets
\citep{sec9:Buc81}. A number of transfer experiments
\citep{sec9:Sch86, sec9:Vog96} providing complementary
information about the threshold levels in $^{27}$Si not accessible
to direct study by radiative capture measurements lead to an
improved reaction rate for $^{26}$Al ground state capture. A first
direct study of a lower energy resonance was successfully
performed in inverse kinematics at the ISAC facility at TRIUMF
using the DRAGON recoil separator \citep{sec9:Rui06}. The
resonance value is substantially smaller than the value quoted
before \citep{sec9:Vog96}, which reduced the reaction rate
slightly at temperatures anticipated for nova burning conditions.

Not included in the reaction rate calculations are possible
contributions of proton capture on the thermally first excited
state in $^{26}$Al \citep{sec9:Run01}. Recently  number of \index{isotopes!26Al}
indirect measurements have been performed to explore the possible
contribution to the total reaction rate of
$^{26}$Al(p,$\gamma$)$^{27}$Si. Transfer reactions have been used
to populate proton unbound states in $^{27}$Si measuring the
subsequent proton decay to the ground state and the first excited
state in $^{26}$Al \citep{sec9:Dei09}. This approach allows to
determine the branching and the relative strength of the proton
decays for each of the unbound states. This can be used to scale
the reaction rate component for the proton capture on the first
excited state.

Possible lower energy resonance contributions to the proton
capture rates on the ground state \citep{sec9:Lot091} and the
excited state of $^{26}$Al \citep{sec9:Lot092} have been explored by
$\gamma$ spectroscopy techniques probing the proton unbound state
in $^{27}$Si through heavy ion fusion evaporation reactions and
measuring the $\gamma$ decay of proton unbound states. This is a
particular efficient method to explore the levels near the
threshold where proton decay is suppressed by the Coulomb barrier.
The measurements provide critical information about spin and
parity of the observed states but gives only limited information
about the resonance strengths which is primarily determined by the
proton decay strength.

The main reaction for the depletion of $^{60}$Fe in neutron rich \index{isotopes!60Fe}
environments is $^{60}$Fe(n,$\gamma$)$^{61}$Fe. The reaction rate
used for nucleosynthesis simulations was for many years based on
theoretical Hauser Feshbach model predictions. Recently an
experiment has been performed at the FZK Karlsruhe in Germany to
determine the stellar reaction cross section experimentally by
neutron activation with the neutron beam resembling a
quasi-stellar neutron spectrum \citep{sec9:Ube09}. The activated
$^{60}$Fe sample was prepared from PSI beamstop material. The
cross section was determined from the characteristic $^{61}$Fe
$\gamma$ activity relative to the amount of $^{60}$Fe nuclei in
the target material. The latter was determined from the
characteristic $^{60}$Fe $\gamma$ activity of the target sample.
based on this the experimental results suggest a cross section
which is twice as large as standard Hauser Feshbach predictions
suggesting a much more rapid depletion of $^{60}$Fe in neutron
rich environment than previously anticipated. The estimate of the
number of $^{60}$Fe nuclei, however relied on adopting a half-life
of T$_{1/2}$=1.49 Gy \citep{sec9:Kut84}. Recent work suggested
that the half-live is considerably larger model T$_{1/2}$=2.62 Gy
\citep{sec9:Rug09}. This would translate into a considerably
larger amount of $^{60}$Fe particles in the sample, suggesting a
cross section which would be in fair agreement with the Hauser
Feshbach predictions. New independent life time measurements for
$^{60}$Fe are clearly necessary to address this issue and remove
the uncertainty in the interpretation of the radiative capture
data.

\bibliographystyle{spbasic}

%
%
%
%
%
%
%
%
%

\begin{thebibliography}{99.}%

\bibitem[Angulo~et~al.(1999)]{sec9:nacre} C. Angulo, M. Arnould, M. Rayet, et al. Nucl.\ Phys. \textbf{A656} 3 (1999)
\bibitem[Arazi ~et~al.(2006)]{sec9:Ara06} A. Arazi, T. Faestermann, J. O. Fernandez Niello, et al., Phys.\ Rev. \ C \textbf{74} 025802 (2006)
\bibitem[Arnould ~et~al.(2007)]{sec9:fiss2} M. Arnould, S. Goriely, K. Takahashi, Phys.\ Rep.\ \textbf{450}, 97 (2007)
\bibitem[Blatt,~et~al.(1991)]{sec9:BW52} J. M. Blatt, V. F. Weisskopf, {\it Theoretical Nuclear Physics} (Dover, 1991)
\bibitem[Borcea ~et~al.(2003)]{sec9:ntof} C. Borcea, P. Cennini, M. Dahlfors, et al., Nucl.\ Instr.\ Meth. A \textbf{513} 524 (2003)
\bibitem[Buchmann~et~al.(1984)]{sec9:Buc81} L. Buchmann, M. Hilgemeier, A. Krauss, et al., Nucl.\ Phys.\ \textbf{A415} 93 (1984)
\bibitem[Chae~et~al.(2006)]{sec9:102} K. Y. Chae, D. W. Bardayan, J. C. Blackmon, et al. Phys.\ Rev.\ C \textbf{74} 012801 (2006)
\bibitem[Cooperman~et~al.(1977)]{sec9:coo77} E. L. Cooperman, M. H. Shapiro, and H. Winkler, Nucl.\ Phys. \textbf{A284} 163 (1977)
\bibitem[Cooperstein~et~al.(1984)]{sec9:CW84} J. Cooperstein, J. Wambach, Nucl.\ Phys.\ \textbf{A420}, 591 (1984)
\bibitem[Costantini ~et~al.(2009)]{sec9:Co09} H. Costantini, A. Formicola, G. Imbriani, et al., Rep.\ Prog.\ Phys \textbf{72} 086301 (2009)
\bibitem[Couder ~et~al.(2008)]{sec9:Cou08} M. Couder, G.P.A. Berg, J. G\"orres, et al., Nucl.\ Instr.\ Meth.\ A \textbf{587} 35 (2008)
\bibitem[Couture ~et~al.(2008)]{sec9:CBC08} A. Couture, M. Beard, M. Couder, et al, Phys.\ Rev.\ C \textbf{77} 015802 (2008)
\bibitem[Cowan ~et~al.(1991)]{sec9:fiss1} J. J. Cowan, F.-K. Thielemann, J. W. Truran, Phys.\ Rep.\ \textbf{208}, 267 (1991)
\bibitem[D'Auria~et~al.(2004)]{sec9:21Na} J. M. D'Auria, R. E. Azuma, S. Bishop, et al., Phys.\ Rev.\ C \textbf{69} 065803 (2004)
\bibitem[De S\'er\'ville~et~al.(2009)]{sec9:103} De S\'er\'ville N, C Angulo, A. Coc, et al. Phys.\ Rev.\ C \textbf{79} 015801 (2009)
\bibitem[Deibel~et~al.(2009)]{sec9:Dei09} C. M. Deibel. J. A. Clark, R. Lewis, et al., Phys.\ Rev.\ C \textbf{80} 035806 (2009)
\bibitem[Descouvemont(2003)]{sec9:DesBook} P. Descouvemont, \textit{Theoretical Models for Nuclear Astrophysics} (Nova Science Publications, 2003)
\bibitem[Descouvemont ~et~al.(2006)]{sec9:DR06} P. Descouvemont, T. Rauscher, Nucl.\ Phys.\ \textbf{A777}, 137 (2006)
\bibitem[Di~Leva ~et~al.(2009)]{sec9:DiL09} A. Di Leva, L. Gialanella, R. Kunz, et al., Phys.\ Rev.\ Lett. \textbf{102} 232502 (2009)
\bibitem[Elix~et~al.(1979)]{sec9:eli79} K. Elix, H. W. Becker, L. Buchmann, et al., Z.\ Phys.\ A \textbf{293} 261 (1979)
\bibitem[Ericson(1960)]{sec9:Eric60} T. Ericson, Phys.\ Rev.\ Lett.\ \textbf{5}, 430 (1960)
\bibitem[Fuller ~et~al.(1982)]{sec9:weak2} G. M. Fuller, W. A. Fowler, M. J. Newman, Ap.\ J. \textbf{252}, 715 (1982)
\bibitem[G\"orres~et~al.(1998)]{sec9:Goe98} J. G\"orres, J. Mei\ss ner, H. Schatz, et al., Phys.\ Rev.\ Lett. \textbf{80} 2554 (1998)
\bibitem[Gadioli ~et~al.(1992)]{sec9:GH92} E. Gadioli, P. E. Hodgson, \textit{Pre-Equilibrium Nuclear Reactions} (Clarendon Press, Oxford 1992)
\bibitem[Glendenning(2004)]{sec9:Glen} N. K. Glendenning, \textit{Direct Nuclear Reactions} (World Scientific, 2004)
\bibitem[Goriely ~et~al.(2008)]{sec9:thalys} S. Goriely, S. Hilaire, A. J. Koning, Astron.\ Astrophys.\ \textbf{487}, 767 (2008)
\bibitem[Goriely ~et~al.(2009)]{sec9:fissgor}  S. Goriely, S. Hilaire, A. J. Koning, M. Sin, R. Capote, Phys.\ Rev.\ C \textbf{79}, 024612 (2009)
\bibitem[Hauser ~et~al.(1952)]{sec9:HF52} W. Hauser, H. Feshbach, Phys.\ Rev. \textbf{87}, 366 (1952)
\bibitem[Heil ~et~al.(2001)]{sec9:Hei01} M. Heil, R. Reifarth, M. M. Fowler, et al.,Nucl.\ Instr.\ Meth.\ A \textbf{459} 516 (2001)
\bibitem[Hix ~et~al.(1999)]{sec9:HT99} W. R. Hix, F.-K. Thielemann, Ap.\ J. \textbf{511}, 862 (1999)
\bibitem[Holmes~et~al.(1976)]{sec9:HWFZ76} J. A. Holmes, S. E. Woosley, W. A. Fowler, B. A. Zimmerman, At.\ Data Nucl.\ Data Tables \textbf{18}, 305 (1976)
\bibitem[Iliadis(2007)]{sec9:Il07} C. Iliadis, {\it Nuclear Physics of Stars} (Wiley-VCH, 2007)
\bibitem[Iliadis~et~al.(1990)]{sec9:ili90} C. Iliadis, T. Schange, C. Rolfs, et al. Nucl.\ Phys. \textbf{A512} 509 (1990)
\bibitem[Jenkins~et~al.(2004)]{sec9:Jen04}  D. G. Jenkins, C. J. Lister, R. V. F. Janssens, et al., Phys.\ Rev.\ Lett. \textbf{92} 031101 (2004)
\bibitem[Johnson~et~al.(1992)]{sec9:JKKL92} C. W. Johnson, E. Kolbe, S. E. Koonin, K. Langanke, Ap.\ J. \textbf{392}, 320 (1992)
\bibitem[Kim ~et~al.(1987)]{sec9:dc} K. H. Kim, M. H. Park, B. T. Kim, Phys.\ Rev.\ C \textbf{23}, 363 (1987)
\bibitem[Knie ~et~al.(2000)]{sec9:Kni00} K. Knie, T. Faestermann, G. Korschinek, et al., Nucl.\ Instr.\ Meth.\ B \textbf{172} 717 (2000)
\bibitem[Knie~et~al.(2004)]{sec9:Kni04} K. Knie, G. Korschinek, T. Faestermann, et al., Phys.\ Rev.\ Lett. \textbf{93} 171103 (2004)
\bibitem[Kolata~et~al.(1989)]{sec9:twinsol} J. J. Kolata, A. Morsad, X. J. Kong, et al., Nucl.\ Instr.\ Meth.\ B \textbf{40} 503 (1989)
\bibitem[Kutschera ~et~al.(1997)]{sec9:Stei02} W. Kutschera, P. Collon, H. Friedmann, et al., Nucl.\ Instr.\ Meth.\ B \textbf{123} 47 (1997)
\bibitem[Kutschera~et~al.(1984)]{sec9:Kut84} W. Kutschera, P. J. Billquist, D. Frekers, et al., Nucl.\ Inst.\ Meth.\ B \textbf{5} 430 (1984)
\bibitem[Lane ~et~al.(1958)]{sec9:LT58} A. M. Lane, R. G. Thomas, Rev.\ Mod.\ Phys.\ \textbf{30}, 257 (1958)
\bibitem[Lisowski ~et~al.(1990)]{sec9:Lansce} P. W. Lisowski, C. D. Bowman, G. J. Russell, S. A. Wender, Nucl.\ Sci.\ Eng. \textbf{106} 208(1990)
\bibitem[Lotay~et~al.(2009)]{sec9:Lot091} G. Lotay, P. J. Woods, D. Sewryniak, et al., Phys.\ Rev.\ Lett.\ \textbf{102} 162502 (2009)
\bibitem[Lotay~et~al.(2009)]{sec9:Lot092} G. Lotay, P. J. Woods, D. Sewryniak, M. P. Carpenter, R. V. E. Janssens, S. Zhu, Phys.\ Rev.\ C \textbf{80} 055802 (2009)
\bibitem[M\"oller~et~al.(2003)]{sec9:weak1} P. M\"oller, B. Pfeiffer, K.-L. Kratz, Phys.\ Rev.\ C \textbf{67}, 055802 (2003)
\bibitem[Nassar ~et~al.(2005)]{sec9:Nas05} H. Nassar, M. Paul, I. Ahmad, et al., Phys.\ Rev.\ Lett.\ \textbf{94} 092504 (2005)
\bibitem[Nassar ~et~al.(2006)]{sec9:Nas06} H. Nassar, M. Paul, I. Ahmad, et al., Phys.\ Rev.\ Lett.\ \textbf{96} 041102 (2006)
\bibitem[Newton ~et~al.(2007)]{sec9:NICCPU07} J. R. Newton, C. Iliadis, A. E. Champagne, A. Coc, Y. Parpottas, C. Ugalde, Phys.\ Rev.\ C \textbf{75}, 045801 (2007)
\bibitem[Panov ~et~al.(2005)]{sec9:fiss3} I. V. Panov, E. Kolbe, B. Pfeiffer, T. Rauscher, K.-L. Kratz, F.-K. Thielemann, Nucl.\ Phys.\ \textbf{A747}, 633 (2005)
\bibitem[Panov ~et~al.(2010)]{sec9:fissigornew} I. V. Panov, I. Yu.\ Korneev, T. Rauscher, G. Mart\'inez-Pinedo, A. Keli\'c-Heil, N. T. Zinner, F.-K. Thielemann, Astron.\ Astrophys.\ \textbf{513}, A61 (2010); arXiv:0911.2181
\bibitem[Per\"aj\"arvi~et~al.(2000)]{sec9:Per00} K. Per\"aj\"arvi, T. Siiskonen, A. Honkanen, P. Dendooven, et al., Phys.\ Lett.\ B \textbf{492} 1 (2000)
\bibitem[Pieper ~et~al.(2001)]{sec9:PW01} S. C. Pieper, R. B. Wiringa, Ann.\ Rev.\ Nucl.\ Part.\ Phys.\ \textbf{51}, 53 (2001)
\bibitem[Ratynski ~et~al.(1988)]{sec9:Rat88} W. Ratynski, F. K\"appeler, Phys.\ Rev.\ C \textbf{37}, 595 (1988)
\bibitem[Rauscher ~et~al.(1996)]{sec9:RR96} T. Rauscher, G. Raimann, Phys.\ Rev.\ C \textbf{53}, 2496 (1996)
\bibitem[Rauscher ~et~al.(1997)]{sec9:RTK97} T. Rauscher, F.-K. Thielemann, K.-L. Kratz, Phys.\ Rev.\ C \textbf{56}, 1613 (1997)
\bibitem[Rauscher ~et~al.(2000a)]{sec9:RT00} T. Rauscher, F.-K. Thielemann, At.\ Data Nucl.\ Data Tables \textbf{75}, 1 (2000a)
\bibitem[Rauscher ~et~al.(2000b)]{sec9:Rau00} T. Rauscher, F. K. Thielemann, J. G\"orres, M. Wiescher, Nucl.\ Phys. \textbf{A675} 695 (2000b)
\bibitem[Rauscher ~et~al.(2002)]{sec9:rhhw92} T. Rauscher, A. Heger, R. D. Hoffman, S. E. Woosley, Ap.\ J. \textbf{576}, 323 (2002)
\bibitem[Rauscher(2010)]{rauenergies} T. Rauscher, Phys.\ Rev.\ C \textbf{81}, 045807 (2010)
\bibitem[Rauscher(2011)]{raureview} T. Rauscher, Int.\ J. Mod.\ Phys.\ E, in press (2011); arXiv:1010.4283
\bibitem[Robertson~et~al.(2007)]{sec9:Rob08} D. Robertson, C. Schmitt, Ph. Collon, et al., Nucl.\ Instr.\ Meth.\ B \textbf{259} 669 (2007)
\bibitem[Rugel~et~al.(2009)]{sec9:Rug09}  G. Rugel, T. Faestermann, K. Knie, et al., Phys.\ Rev.\ Lett.\ \textbf{103} 072502 (2009)
\bibitem[Ruiz~et~al.(2006)]{sec9:Rui06} C. Ruiz, A. Parikh, J. Jos\'e, L. Buchmann, Jet al., Phys.\ Rev.\ Lett. \textbf{96} 252501 (2006)
\bibitem[Runkle ~et~al.(2005)]{sec9:RCA05} R. C. Runkle, A. E. Champagne, C. Angulo, et al. Phys.\ Rev.\ Lett. \textbf{94:} 082503 (2005)
\bibitem[Runkle~et~al.(2001)]{sec9:Run01} R. C. Runkle, A. E. Champagne, J. Engel, Ap\ J \textbf{556}, 970 (2001)
\bibitem[Salpeter~et~al.(1969)]{sec9:SvH69} E. E. Salpeter, H. M. Van Horn, Ap.\ J. \textbf{155}, 183 (1969)
\bibitem[Satchler(1983)]{sec9:Sat} G. R. Satchler, \textit{Direct Nuclear Reactions} (Clarendon, Oxford 1983)
\bibitem[Schmalbrock~et~al.(1986)]{sec9:Sch86} P. Schmalbrock, T. R. Donoghue, M. Wiescher, V. Wijekumar, C. P. Browne, A. A. Rollefson, C. Rolfs, A. Vlieks, Nucl.\ Phys. \textbf{A457}, 182 (1986)
\bibitem[Seuthe~et~al.(1990)]{sec9:Seu89} S. Seuthe, C. Rolfs, U. Schr\"oder, et al., Nucl.\ Phys.\ \textbf{A514} 471 (1990)
\bibitem[Simpson~et~al.(1971)]{sec9:di71} J. J. Simpson, W. R. Dixon, and R. S. Storey, Phys.\ Rev.\ C \textbf{4} 443 (1971)
\bibitem[Stegm\"uller~et~al.(1996)]{sec9:Ste96} F. Stegm\"uller, C. Rolfs, S. Schmidt, et al., Nucl.\ Phys.\ \textbf{A601} 168 (1996)
\bibitem[Terrasi ~et~al.(2007)]{sec9:Ter07} F. Terrasi, D. Rogalla, N. De Cesare, et al., Nucl.\ Instr.\ Meth.\ B \textbf{259} 14 (2007)
\bibitem[Uberseder~et~al.(2009)]{sec9:Ube09} E. Uberseder, R. Reifarth, D. Schumann, et al., Phys.\ Rev.\ Lett.\ \textbf{102} 151101 (2009)
\bibitem[Vockenhuber ~et~al.(2007)]{sec9:Voc07} C. Vockenhuber, C. O. Ouellet, L.-S. The, et al.,  Phys.\ Rev.\ C \textbf{76} 035801 (2007)
\bibitem[Vogel(2006)]{sec9:weak3} P. Vogel, Nucl.\ Phys.\ \textbf{A777}, 340 (2006)
\bibitem[Vogelaar~et~al.(1996)]{sec9:Vog96} B. Vogelaar, L. W. Mitchell. R. W. Kavanagh, et al., Phys.\ Rev. \ C \textbf{53} 1945 (1996)
\bibitem[Wagoner ~et~al.(1969)]{sec9:Wag69} R. V. Wagoner, Ap.\ J. Suppl.\ \textbf{18}, 247 (1969)
\bibitem[Ward~et~al.(1980)]{sec9:WF80} R. A. Ward, W. A. Fowler, Ap.\ J. \textbf{238}, 266 (1980)
\bibitem[Yakovlev~et~al.(2006)]{sec9:YGA06} D. G. Yakovlev, L. R. Gasques, A. V. Afanasjev, et al., Phys.\ Rev.\ C \textbf{74}, 035803 (2006)


\end{thebibliography}


\end{document}